\documentclass[10pt,journal,compsoc]{IEEEtran}
\usepackage{booktabs} 
\usepackage{tikz}

\usepackage{bm}
\usepackage{multirow}
\usepackage{multicol}
\usepackage{hhline}
\usepackage{amsmath}
\usepackage{subcaption}
\usepackage{cases}
\usepackage{pgfplots}
\usepackage{bm}

\usepackage{color}
\usepackage{xcolor}
\usepackage{amssymb}
\usepackage{amsfonts}
\usepackage{ragged2e}
\usepackage[utf8]{inputenc}
\usepackage{csvsimple}
\usepackage[T1]{fontenc}

\usepackage{amsmath,amssymb,amsfonts}
\usepackage{algorithmic}
\usepackage{graphicx}
\usepackage{textcomp}
\usepackage{xcolor}
\pgfplotsset{compat=1.14}
\def\BibTeX{{\rm B\kern-.05em{\sc i\kern-.025em b}\kern-.08em
    T\kern-.1667em\lower.7ex\hbox{E}\kern-.125emX}}

\ifCLASSOPTIONcompsoc
  \usepackage[nocompress]{cite}
\else
  \usepackage{cite}
\fi
\ifCLASSINFOpdf
\else
\fi
\begin{document}
%
\title{On the Performance Analysis of Epidemic Routing in Non-Sparse Delay Tolerant Networks}
%
%
%
%

\author{Leila~Rashidi,
        Don~Towsley,~\IEEEmembership{Fellow,~IEEE},
        Arman Mohseni-Kabir,
        and~Ali~Movaghar,~\IEEEmembership{Senior~Member,~IEEE}
\IEEEcompsocitemizethanks{\IEEEcompsocthanksitem L. Rashidi is with the Department of Computer Science, University of Calgary, Calgary, AB, Canada and the College of Information and Computer Sciences, University of Massachusetts, Amherst, MA, USA.\protect\\
E-mail: leila.rashidi@ucalgary.ca
\IEEEcompsocthanksitem D. Towsley is with the College of Information and Computer Sciences, University of Massachusetts, Amherst, MA, USA.\protect\\
E-mail: towsley@cs.umass.edu

\IEEEcompsocthanksitem A. Mohseni-Kabir is with the Department of Physics, University of Massachusetts, Amherst, MA, USA.\protect\\
E-mail: arman@physics.umass.edu

\IEEEcompsocthanksitem A. Movaghar is with the Department of Computer Engineering, Sharif University of Technology, Tehran, Iran.\protect\\
E-mail: movaghar@sharif.edu

}
\thanks{The work was done when the first author was a visiting student in the College of Information and Computer Sciences, University of Massachusetts under the supervision of the second author.}}

\IEEEtitleabstractindextext{%
\begin{abstract}
We study the behavior of epidemic routing in a delay tolerant network as a function of node density. Focusing on the probability of successful delivery to a destination within a deadline (PS), we show that PS experiences a phase transition as node density increases. Specifically, we prove that PS exhibits a phase transition when nodes are placed according to a Poisson process and allowed to move according to independent and identical processes with limited speed. We then propose four fluid models to evaluate the performance of epidemic routing in non-sparse networks. A model is proposed for supercritical networks based on approximation of the infection rate as a function of time. Other models are based on the approximation of the {\em pairwise infection rate}. Two of them, one for subcritical networks and another for supercritical networks, use the pairwise infection rate as a function of the number of infected nodes. The other model uses pairwise infection rate as a function of time, and can be applied for both subcritical and supercritical networks achieving good accuracy. The model for subcritical networks is accurate when density is not close to the percolation critical density. Moreover, the models that target only supercritical regime are accurate.
\end{abstract}

}

\maketitle

\IEEEdisplaynontitleabstractindextext

%
\IEEEpeerreviewmaketitle

\IEEEraisesectionheading{\section{Introduction}\label{sec:intro}}

%
%
%
%
\IEEEPARstart{A}{} delay tolerant network (DTN) is a wireless network that provides communication between mobile nodes based on the store-carry-forward paradigm \cite{cao2013routing, 8316884}. DTNs have many applications such as mobile data offloading, opportunistic content sharing, opportunistic experience sharing, and satellite communication \cite{6517187, 7524493, 7994721}. 
There are scenarios in the real world, where many mobile nodes are connected \cite{Heimlicher}. This can occur in places such as stadiums, shopping malls, airports, and cafeterias where opportunistic contacts can be used to transfer delay tolerant traffic to decrease the load on infrastructure networks~\cite{KARKKAINEN201865}. When a wireless network is not sparse, large clusters of nodes can form wherein nodes are connected to each other. One evidence of non-sparse networks is the RollerNet trace \cite{5062024} that exhibits a dynamic dense network with high connectivity \cite{5282448}. Furthermore,
\cite{Pietilanen:2012:DOS:2248371.2248396} studies four real-life human mobility traces, namely \textit{INFOCOM} \cite{cambridge-haggle-20090529}, \textit{SIGCOMM} \cite{thlab-sigcomm2009-20120715}, \textit{Reality} \cite{mit-reality-20050701}, and \textit{Strathclyde} \cite{strath-nodobo-20110323}, and discovers a large cluster whose membership varies over time. The store-carry-forward paradigm can be used for communication in non-sparse mobile networks. Even in dense networks, if node mobility is high, mobile ad-hoc network (MANET) routing schemes cannot be used to provide communication between a subset of nodes since clusters are so dynamic, and paths are momentary.


Non-sparse wireless networks have attracted attention from the research community from both theoretical and practical perspectives \cite{gupta1999critical, 4106120, Kong2008, 5282448, Heimlicher, KARKKAINEN201865, 8057301}. Several studies have shown the existence of phase transitions in such networks \mbox{\cite{gupta1999critical, haenggi2009stochastic}}. A phase transition is defined as a sudden radical change in a spatial process that occurs as a parameter changes in a continuous manner \cite{continuumPercolation}. Phase transition phenomena can be exploited to design more efficient networks \cite{Hui:2008:PTO:1409985.1409999}. Often when a phase transition occurs as a consequence of changing a parameter such as node density, setting the parameter just below or above the critical point can yield much better performance. Gilbert's random disk graph is model of a static wireless network. The random disk graph nodes are placed in a 2-D Euclidean space according to a Poisson point process $\Phi$ with density $\lambda$. A random disk graph with range $R$ includes an edge between two nodes if their distance is not more than $R$ \cite{haenggi2009stochastic}. Let a cluster of nodes denote a set of nodes such that each pair of nodes are connected directly via an edge or through a path of nodes. The probability of existence of an infinite cluster, i.e., a cluster containing an infinite number of nodes, in Gilbert's random disk graph experiences a phase transition as node density increases~\cite{haenggi2009stochastic}. \textit{Percolation} refers to the property of having an infinite cluster \cite{haenggi2009stochastic}.

\cite{Kong2008} shows that the delivery delay of epidemic routing scales linearly with the initial distance between the source and destination nodes in a subcritical network, i.e., a network whose density is less than a critical density associated with the existence of an infinite cluster. On the other hand, it scales sublinearly with the initial distance in a supercritical network, i.e., a network whose density is more than the critical density \cite{Kong2008}. Based on these results, we believe that performance of epidemic routing may exhibit a phase transition at the critical density associated with the existence of an infinite cluster. 


Fluid models (based on ordinary differential equations (ODEs)) and Markov chains are the two main approaches used to study the performance of DTNs. These models have been successfully used to evaluate the performance of routing schemes in sparse networks. To the best of our knowledge, previous ODE and Markov models proposed to evaluate the performance of DTNs are based on the assumption that clusters are typically at most of size two, which is reasonable for a sparse network. These models primarily focus on modeling meetings between one infected node and one uninfected node. When such a meeting occurs, the uninfected node becomes infected after a short delay, which is usually assumed to be zero in the literature \cite{zhang2007performance, picu2015dtn, wang2015restricted}. Thus, the number of infected nodes increases by one.
This modeling approach does not account for the possibility of large clusters that form in non-sparse networks. In the case of a non-sparse network, when an infected node meets a node in an uninfected large cluster, the message transfers to all nodes of the cluster quickly, and more than one node is infected in a small time interval. This raises the question: how does one evaluate the performance of epidemic routing in a non-sparse DTN? Non-sparsity of DTNs impacts network performance in two opposing ways. First, the connectivity of a large cluster enables fast propagation of a recently received message inside a cluster. This has a positive effect on performance in terms of delivery delay. Second, the speed of transmissions is reduced due to interference and the need to share the bandwidth and scheduling transmissions. In this paper, we focus on the performance of epidemic routing in a non-sparse DTN considering the first (positive) effect. Indeed, considering the second effect is challenging, and we leave it for future work.

Measures of interest include the average number of infected nodes at any time and probability of successful delivery within a deadline (PS), which corresponds to the ratio of the average number of infected
nodes to total number of nodes excluding the source. It is worth mentioning that the average delivery delay of epidemic routing in the case of no deadline follows in a straightforward manner after the average number of infected nodes as a function of time is computed. 
According to epidemic routing, each infected node forwards the message to any node within transmission range, but nodes are restricted in forwarding the message under other routing schemes such as spray and focus and $K$-hop forwarding. Thus, if the network is lightly loaded such that nodes have enough free space in their buffer, epidemic routing achieves the highest PS. Epidemic routing may not be efficient due to its large communication cost when nodes have space/energy constraints. However,
studying
performance of epidemic routing provides good insights in terms of lower or upper
bounds for performance measures such as average delivery delay and average number of infected nodes in the case of using other routing schemes. These results can be used for design, evaluation, and optimization
of the routing schemes. Epidemic routing has been extensively considered by the research community in recent years \cite{7103308, yang2016delay, sermpezis2016delay, 7473919, 7858750, 8681155}.


First, we prove that PS of epidemic routing exhibits a phase transition as node density increases in an infinite network. We study the PS of epidemic routing in large finite networks using simulation. Simulation results indicate that a phase transition happens in PS of epidemic routing as node density or transmission range increases. In addition to studying the performance of finite networks using simulation, we investigate it from an analytical perspective. To this end, we first discuss shortcomings of previous ODE models in the performance analysis of non-sparse networks. This motivates four new ODE models for epidemic routing in non-sparse networks. Three models are designed based on the notion of \textit{pairwise infection rate} defined as the infection rate per pair of nodes that includes one infected node and one uninfected node. They differ from each other according to how the pairwise infection rate is approximated. In two models, we approximate the pairwise infection rate as a function of the number of infected nodes while in the other models, we approximate the infection rate and the pairwise infection rate as functions of time. These approximations are developed using results obtained from discrete-event simulation. The model using the pairwise infection rate as a function of time can be applied to evaluate the performance of epidemic routing in both subcritical and supercritical networks. However, the other models can be used for either subcritical networks or supercritical networks.

We verify the occurrence of phase transitions and validate the proposed ODE models through discrete-event simulation of non-sparse DTNs. Results obtained from the simulation show that PS of a large finite network exhibits a behavior similar to the phase transition in an infinite network. Comparison of analytic results of PS, computed from the proposed model for subcritical networks that uses the pairwise infection rate as a function of the number of infected node, with results obtained from simulation indicates that this model is very accurate when the density is not close to the percolation critical density, but the accuracy of this model decreases as the density increases. The results computed from the proposed models for supercritical networks are close to results obtained from simulation. Moreover, the proposed model using the pairwise infection rate as a function of time yields good accuracy to evaluate the performance of epidemic routing in both subcritical and supercritical regimes. In order
to show the superiority of the proposed models, we compare them
with previous ODE model proposed in \cite{zhang2007performance}. All proposed models
outperform the previous ODE model.


The main contributions of this paper are
\begin{itemize}
\item We prove that PS of epidemic routing experiences a phase transition at a critical density that is the same as the percolation critical density for the existence of an infinite cluster when nodes move according to independent identically distributed (i.i.d.) stochastic processes with limited speed. 

\item We find that the pairwise infection rate as a function of time, exhibits exponential decay, and we model it as such. On the other hand, we find that the pairwise infection rate of a subcritical or supercritical network as a function of the number of infected nodes, exhibits power law behavior or exponential decay, respectively. Moreover, we show that the infection rate in supercritical networks decays exponentially.
\item Comparing results computed from the standard ODE model \cite{zhang2007performance} with results obtained from the simulation, we show that this model is not accurate when used to model a non-sparse network.
Afterwards, we propose four ODE models for the analysis of epidemic routing in non-sparse DTNs. We obtain closed form solutions for the models which use the pairwise infection rate and the infection rate as a function of time. Using these solutions, we derive two expressions for the average number of infected nodes at any time under epidemic routing.
\end{itemize}

This paper is organized as follows. First, a literature review is presented in Section~\ref{sec:related_work}. Section~\ref{sec:discreption} is dedicated to explain the target networks and assumptions. We argue phase transition behavior of PS in Section~\ref{sec:transition}. Afterwards, in Section~\ref{sec:pairwiseRate}, we propose three ODE models for the epidemic routing in non-sparse networks. Section~\ref{sec:perf} presents numerical results. Finally, Section~\ref{sec:conc} concludes the paper, and presents future research directions.

\section{Related Work} \label{sec:related_work}
In this section, we first introduce the literature. Afterwards, we compare our work with previous approaches in Section~\ref{sec:related_work_com}, and show how we move the current research forward.
Phase transitions in wireless ad-hoc networks have been studied in several papers. 
\cite{gupta1999critical} proves that a phase transition happens in connectivity of an ad-hoc network when all nodes have a fixed communication range. \cite{965963} shows that when a sensor tracking network is modeled as a Bernoulli random graph model, a phase transition occurs in the probability of tracking all targets as the edge probability increases. Furthermore, ~\cite{965963} shows that the probability of delivering a message to all nodes using probabilistic flooding exhibits a phase transition as the forwarding probability increases. \cite{965963} assumes that the source forwards the message whenever it meets a node. \cite{sasson2003probabilistic} shows that in an ideal MANET, a phase transition is observed in the average ratio of distinct messages delivered to each node by the total number of distinct messages which are flooded using probabilistic flooding. Furthermore, \cite{sasson2003probabilistic} shows, using simulation, that this ratio does not exhibit a phase transition when realistic effects such as low node mobility and packet collisions are accounted for. 

\cite{6151313} studies a linear network, i.e., nodes are located in a row, where there exists a link between any two successive nodes. At any time instant, each link is in either the down state or the up state, and link dynamics are governed by mutually independent identically distributed stochastic processes. Considering the first and the last nodes as the source and the destination, respectively, \cite{6151313} proves that the probability of
successful delivery within a deadline exhibits a phase transition as deadline increases. \cite{7041091} shows that the fraction of nodes that receive a given message and the ratio of total number of transmissions of messages and total number of generated messages exhibit phase transitions under probabilistic flooding. These phase transitions occur when the forwarding probability varies
from 0 to 0.1.
\cite{mostafizi2017percolation} studies phase transition phenomenon in a connected vehicle network, and shows that a phase transition occurs in the average travel time of vehicles as the communication range of a vehicle increases.

\cite{6562898} studies the effect of accounting for temporary multi-hop paths in making forwarding decisions on the performance of WAIT forwarding strategy. \cite{6562898} suggests that monitoring a node's vicinity up to four hops improves network performance while monitoring overhead is low. Subsequently, \cite{6814723} studies the predictability of multi-hop opportunities in DTNs. The performance of non-sparse wireless networks has been studied from a percolation theory perspective in \cite{4106120, shen2006directional, Kong2008, kong2009connectivity,  Zhao:2011:FRN:2030613.2030651, peres2013mobile, 6568976}. \cite{4106120} and \cite{shen2006directional} study a wireless network with static nodes. In \cite{4106120}, percolation theory was exploited to derive information capacity results for random networks. In \cite{shen2006directional}, probabilistic broadcast in two cases of using omnidirectional antennas and directional antennas was mapped to site percolation and bond percolation, respectively. Subsequently, these mappings were used to compare the performances of probabilistic broadcast in the aforementioned two cases.

\cite{Kong2008} assumes that nodes are initially placed according to a Poisson point process in $\mathbb{R}^2$ with density~$\lambda$, and that they move according to a constrained i.i.d. mobility model. \cite{Kong2008} proves that the delivery delay of epidemic routing scales linearly with the initial Euclidean distance between the source and the destination if $\lambda^{l}_{c}<\lambda<\lambda^{p}_{c}$ while it scales sublinearly with the initial distance if $\lambda>\lambda^{p}_{c}$ where $\lambda^{l}_{c}$ and $\lambda^{p}_{c}$ are critical densities for long-term connectivity and percolation, respectively. 
\cite{kong2009connectivity} assumes that static nodes are placed according to Poisson point process in $\mathbb{R}^2$, and a link exists between any two nodes located within transmission range of each other, which is active according to a Markov on-off process. \cite{kong2009connectivity} proves that the probability of existence of an infinite cluster of nodes in the aforementioned network model experiences a phase transition as the density of nodes increases. Moreover, \cite{kong2009connectivity} shows that the subcritical and supercritical regimes exhibit similar delivery delay behaviors as \cite{Kong2008}.

\cite{Zhao:2011:FRN:2030613.2030651} and \cite{6568976} model the network as a discrete-time random connection model where at each time, there exists a link between any two nodes that are located in each other's transmission range with a probability that depends on their distance. \cite{Zhao:2011:FRN:2030613.2030651} and \cite{6568976} study the behavior of the ratio of delay and distance as a function of density. They establish that this function is uniformly bounded and monotone decreasing. Moreover, this function is zero if density is more than the critical density for instantaneous. They obtain lower and upper bounds for the ratio of delay and distance. In \cite{peres2013mobile}, three fundamental measures detection, coverage, and percolation time were studied under the assumption that nodes are initially placed according to a Poisson point process in $\mathbb{R}^d$, and move according to a discrete-time brownian motion model. It also presents an upper bound on the time until all nodes have received a message with broadcast in a finite network.

Extensive research has focused on development of analytic models to evaluate the performance of different routing schemes. ODE and Markovian models are two main approaches to analyze the performance of DTNs. \cite{zhang2007performance} and \cite{groenevelt2005message} are early works proposing ODEs and Markovian models, respectively. The ODE approach has been extensively applied in the literature \cite{zhang2007performance, banerjee2008relays, ip2008performance, spyropoulos2009routing, hernandez2017analytical}. Many models have been proposed for epidemic-based communication methods \cite{8606136}. In \cite{ip2008performance}, a continuous time Markov chain was proposed for epidemic routing in a heterogeneous network with two classes of nodes. \cite{chaintreau2009age} uses a system of partial differential equations to study an opportunistic
content update system. \cite{wang2015restricted}, \cite{diana2012modelling}, and \cite{8057110} apply Markov chains to model restricted epidemic routing schemes, spray and wait scheme, and contact avoidance routing, respectively. Furthermore, epidemic routing was studied in \cite{sermpezis2016delay}, \cite{Jindal}, and \cite{5719289}. Some approximations for average delay under epidemic routing were derived in~\cite{sermpezis2016delay}. Delay of epidemic routing was analyzed in~\cite{Jindal} while taking contention into account.
\cite{5719289} proposes an edge-Markovian dynamic graph model for epidemic routing. It is worth mentioning that epidemic content retrieval scheme in sparse DTNs with restricted mobility was modeled in~\cite{8681155} as two monolithic and folded stochastic reward nets.


\subsection{Comparison}\label{sec:related_work_com}
Unlike \cite{965963, 4106120, shen2006directional, kong2009connectivity, Zhao:2011:FRN:2030613.2030651, 6568976} which assume that nodes are static, we study mobile networks. Moreover, unlike \cite{Zhao:2011:FRN:2030613.2030651, peres2013mobile, 6568976, 6151313} which assume time is discrete, we consider continuous time. In this paper, we first study the phase transition phenomenon in PS of epidemic routing. The closest approaches to this study are \cite{965963, sasson2003probabilistic, Kong2008, 6151313, 7041091}. Although we are inspired by \cite{Kong2008}, we prove the occurrence of a phase transition in PS while \cite{Kong2008} studies scaling behavior of the delivery delay. \cite{965963} and \cite{sasson2003probabilistic} discuss the occurrence of phase transition in probability of delivering a message to all nodes and the average fraction of messages delivered to each node, respectively, under probabilistic flooding in MANETs while we focus on the probability of successful delivery within a deadline, PS, under epidemic routing in DTNs. Furthermore, \cite{7041091} studies probabilistic flooding. \cite{6151313} studies the behavior of PS as a function of deadline in a linear random dynamic network while we consider DTNs, and focus on the behavior of PS as a function of node density. 

We also propose four ODE models for epidemic routing in non-sparse DTNs.
The main difference of our approach with previous approaches \cite{groenevelt2005message, Jindal, zhang2007performance, banerjee2008relays, ip2008performance, spyropoulos2009routing, chaintreau2009age, diana2012modelling, picu2015dtn, wang2015restricted, 8057110, sermpezis2016delay, hernandez2017analytical, 8681155} is that we target non-sparse networks in this paper while the aforementioned approaches study sparse networks wherein large clusters of nodes do not exist. In \cite{sermpezis2016delay}, \textit{dense network} refers to a contact network wherein each pair of nodes meet with non-zero probability. 
It is worth mentioning that \cite{5719289} assumes time is discrete which is an oversimplification. Moreover, the state of each edge in the model proposed in \cite{5719289} changes independently from states of other edges. This is not realistic.






\section{Network Model} \label{sec:discreption}
In this paper, we target both finite and infinite networks. Let $\lambda$ denote node density. For the finite case, we consider an~$L\times L$ square containing \mbox{$M=\lambda\cdot L^2$} nodes placed uniformly at random within the square ($M\in \mathbb{N}$). Like \cite{wang2015restricted} and \cite{8681155}, we assume that nodes move according to the random direction mobility model with reflection at the boundaries~\cite{nain2005properties}, where speed and travel length are chosen uniformly from intervals $(v_{min},v_{max}]$ and $(0,t_{r})$ ($0<v_{max},~t_{r}<\infty$), respectively.

Infinite networks have attracted attention from the network research community in the past years \cite{gupta1999critical, Kong2008, kong2009connectivity, Zhao:2011:FRN:2030613.2030651, peres2013mobile, 6568976,hyytia2014searching, sermpezis2016delay}. We consider an infinite \mbox{2-D} plane where mobile nodes are
initially placed according to a two-dimension homogeneous
Poisson point process with density $\lambda$ as \cite{Kong2008}. Nodes move according
to i.i.d. stochastic processes such that the speed of each node is
always less than or equal to $v_{max}$. Thus, node placement at any time~$t$ is a Poisson point process \cite{VANDENBERG1997247}. Therefore, this infinite network at any time $t$ can be considered as a random geometric graph $G_t$, and the probability of existence of an infinite cluster at time $t$ experiences a phase transition as node density increases. Similarly, node placement at time $t$ in a finite network can be considered as a random graph.
Large finite random graphs exhibit a behavior similar to percolation [17]. 
Thus, the probability of existence of a cluster containing a large fraction of nodes at an arbitrary time $t$ in a large finite network exhibits a phase transition as node density increases. 

There is no infrastructure such as Wi-Fi access points in the network, and the network operates on the basis of transmission of messages using short-range communication technologies during opportunistic contacts. Nodes are assumed to transmit to each other whenever they are located in each other's transmission range, which is fixed and denoted by $R$. As mentioned in Section \ref{sec:intro}, we do not study the negative effect of high node density on the performance of network due to interference and scheduling transmissions. Thus, we assume that no interference occurs while transmissions are not scheduled. 
This assumption is also enforced in simulations.

In this paper, we focus on the performance of an epidemic routing scheme where one of the nodes, denoted \textit{source}, wants to deliver a message to another node, denoted \textit{destination}. Both source and destination are chosen uniformly from the mobile nodes. We consider a scenario where the message has to be delivered within a finite time~$T$. This can arise in time-sensitive applications such as opportunistic emergency support systems or content update systems where the content becomes outdated and useless after some time. It can be easily implemented by setting the lifetime of the message to deadline $T$. The measures of interest are the probability of
successful delivery (PS) within deadline $T$, denoted by~$p(T)$, and the average number of infected nodes, i.e., nodes that have the message, at any time~$t$, denoted by $\overline{N}(t)$. Furthermore, we focus on the delivery of only one message as \cite{6151313, wang2015restricted, 8681155, 8843121}, and assume that all nodes have sufficient buffer capacity to store the message \cite{zhang2007performance, boldrini2014performance, picu2015dtn, wang2015restricted, 7473919, 8681155, 8843121}. However, when the total size of all messages that are simultaneously propagated in the network does not exceed the buffer capacity of each node, our approach works, and our results apply to this case.





\section{Phase transition in the probability of success}\label{sec:transition}
In this section, we show that under epidemic routing, $p(T)$ experiences a phase transition. As mentioned in Section~1, a phase transition corresponds to a sudden  radical  change  in  a spatial  process  occurring  as  a parameter  changes  in  a  continuous  manner \cite{continuumPercolation}. For example, network connectivity and the existence of an infinite cluster experience a phase transition as the transmission range or node density changes in a random geometric graph \cite{penrose2003random}, which is one model of continuum percolation. The critical transmission ranges and densities for such transitions have been previously studied \cite{gupta1999critical, quintanilla2000efficient, kong2007characterization, continuumPercolation} through simulation or by analytically computing lower and upper bounds.

We focus on $p(T)$, the probability of successful delivery by time $T$, which is an important measure of performance for a wireless network, and investigate whether it exhibits a phase transition. To simplify analysis, we first assume that the time to transfer a message from one node to another when within distance $R$ of each other is zero, and then we consider positive transfer delays. This assumption has been extensively used in the literature \cite{groenevelt2005message, zhang2007performance, Kong2008, Zhao:2011:FRN:2030613.2030651, 6151313, peres2013mobile, picu2015dtn, wang2015restricted, 7473919, 8681155, 8843121}. It is reasonable because transfer delays are typically much smaller than meeting durations and inter-meeting times, which are the times until infected nodes meet uninfected nodes \cite{Zhao:2011:FRN:2030613.2030651}. Under the aforementioned assumption, whenever an infected node meets an uninfected node, all nodes belonging to the cluster to which the uninfected node belongs immediately become infected.

\begin{figure}
\centering
\begin{tikzpicture}
\begin{axis}[
    xlabel={Node Density, $\lambda$ (nodes/$km^{2}$)},
    ylabel={The Probability of Success, $p(T)$},
    xmin=400, xmax=650,
    ymin=0, ymax=1,
   xtick={400,450,500 ,550, 600, 650},
    ytick={0,0.2,0.4,0.6,0.8,1},
    legend pos=north west,
    ymajorgrids=true,
    grid style=dashed,
]

    \addplot[
    color=gray,
    mark=circle,
    mark options={solid},
    style=dashed,
    forget plot
   ]
    coordinates {
        (565,	0)
        (565,	1)
    };\label{plot_gray}

          \addplot[
    color=gray,
   mark=circle,
    mark options={solid},
    style=dashed,
    forget plot
    ]%
    coordinates {
        (572,	0)
        (572,	1)
    };\label{plot_gray}

    \addplot[
    color=blue,
    mark=square,
    mark options={solid},
    style={thick},
    mark size=1pt
   ]
    coordinates {

         (400,0.0090)
        (425,0.0140)
        (450,0.0250)
        (475,0.0490	)
        (500,0.1096)
        (525,0.3001)
        (550,0.6030) 
        (565, 0.790371)
        (572, 0.821139203)
        (575,0.836824081)
        (600,0.9217)
        (625,0.9522)
       (650,0.9696) 
        
    };\label{plot_blue}

   \addplot[
    color=red,
    mark=triangle,
    mark options={solid},
     style={thick},
    mark size=1pt
   ]
    coordinates {
          (400,0.002289566)
        (425,0.003749448)
        (450,0.0068584)
        (475,0.014351577	)
        (500,0.037876234)
        (525,0.134175847)
        (550,0.51702273)
        (565, 0.776633618)
       (572,0.842318647)
        (575,0.864590832	)
        (600,0.937110096)
        (625,0.960345067)
       (650,0.973348709) 
        
    };\label{plot_red}
    
    \addplot[
    color=green,
    mark=diamond,
    mark options={solid},
     style={thick},
    mark size=1pt
    ]
   coordinates {
        (400,0.001052315)
        (425,0.001709972)
        (450,0.003086774)
        (475, 0.006765583	)
        (500,0.018650502)
        (525,0.075647981)
        (550,0.443303126)
         (565, 0.769263642)
       (572,0.856539738)
        (575,0.878018744	) 
        (600,0.945934061)
        (625,0.96438411)
       (650,0.976455839) 
        
    };\label{plot_green}

        \addplot[
    color=purple,
    mark=*,
    mark options={solid},
     style={thick},
     mark size=1pt
    ]
   coordinates {
        (400,0.000597616)
        (425,0.000969676)
        (450,0.00176459)
        (475,0.003909355)
        (500,0.011054894	)
        (525,0.047970648	)
        (550,0.358969765)
        (565, 0.759883701)
        (572, 0.859015682)
        (575,0.908375254	) 
        (600,0.954090309	)
        (625,0.964718194)
       (650, 0.9755) 
        
    };\label{plot_purple}
    
     \addlegendimage{/pgfplots/refstyle=plot_blue}\addlegendentry{$L$ = 5~$km$}
     \addlegendimage{/pgfplots/refstyle=plot_red}\addlegendentry{$L$ = 10~$km$}
      \addlegendimage{/pgfplots/refstyle=plot_green}\addlegendentry{$L$ = 15~$km$}
     \addlegendimage{/pgfplots/refstyle=plot_purple}\addlegendentry{$L$ = 20~$km$}



\end{axis}
\end{tikzpicture}
\caption{The probability of success for $T=25~s$, $R=50~m$, $v_{min}=0$, $v_{max}=1~m/s$, and $t_{r}=120~s$. Vertical lines represent $\lambda=565$ and $\lambda=572$.
}
\label{fig:transition}
\end{figure}
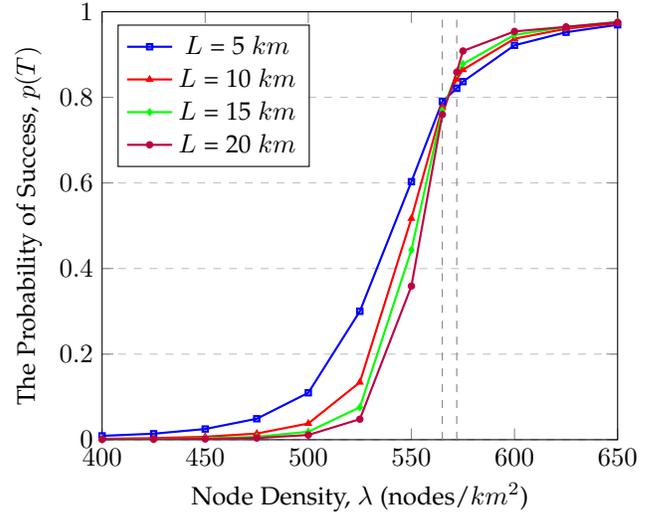

\figurename~\ref{fig:transition} presents $p(T)$ under epidemic routing, obtained from a simulation of a finite network where node density varies from 400 to 650~nodes$/km^{2}$, and the deadline, $T$, is 25~$s$. 
We observe in Fig.~\ref{fig:transition} the appearance of a phase transition in $p(T)$. Let $\lambda_{c}$ denote the percolation critical density, i.e., the critical density at which the phase transition associated with the appearance of an infinite cluster occurs. Simulation studies indicate that $1.43 \leq \lambda_c \leq 1.44$ with more than $99.99\%$ confidence when the transmission range of each node, $R$, is 1 \cite{Kong2008, quintanilla2000efficient}. Thus, the lower and upper bounds represented in (\ref{eq:threshold}) can be estimated for the percolation critical density in the general case.
\begin{equation}\label{eq:threshold}
  \frac{1.43}{R^2} \lessapprox \lambda_c \lessapprox \frac{1.44}{R^2}
\end{equation} According to (\ref{eq:threshold}), 572~nodes/$km^{2}\lessapprox\lambda_{c}\lessapprox 576$ nodes/$km^{2}$ when \mbox{$R=50~m$}. As $L$ increases, the curves for different values of $L$ in Fig.~\ref{fig:transition} intersect around $\lambda=572$. Moreover, the transition of $p(T)$ becomes sharper and approaches a vertical line in the interval $(565,572)$ as $L$ increases. Thus, we expect that $p(T)$ experiences a phase transition when $L=\infty$ around the percolation critical density.
In the following, we prove that under epidemic routing, $p(T)$ experiences a phase transition at the percolation critical density in an infinite network (described in Section~\ref{sec:discreption}).

\newtheorem{lemma}{Lemma}
\begin{lemma} \label{lemma:1}
Let $N(t)$ denote the number of infected nodes at time $t$. For $\lambda$ less than the percolation critical density, \mbox{($\lambda<\lambda_{c}$)}, $\exists~t_0>0$ such that $N(t)$ is finite for $t\in (0,t_0]$ almost surely (a.s.).
\end{lemma}
\begin{IEEEproof}
Every node moves at most a distance $t\cdot v_{max}$ in the interval~$[0,t]$. Thus, any two nodes with initial distance greater than $2\cdot t\cdot v_{max}+R$ do not meet each other in the interval~$[0,t]$. As a result, the initial distance between any two nodes such that one is infected by the other in the interval~$[0,t]$ must be less than or equal to $2\cdot t\cdot v_{max}+R$. Now, we construct a new network with density $\lambda$ that has the same number of nodes as the original network, and the communication range of nodes is $2\cdot t\cdot v_{max}+R$. There is a one-to-one mapping between nodes of this network and those of the original network. The initial position of each node in the new network is the same as the initial position of the corresponding node in the original network.

As mentioned above, in the original network, the initial distance of any two nodes such that one of them is infected by the other in the interval~$[0,t]$ must be less than or equal to $2\cdot t\cdot v_{max}+R$.
Therefore, in the constructed network, the cluster to which the node corresponding to the source belongs at time~0 contains all nodes that correspond to nodes of the original network that are infected in the interval~$[0,t]$. Thus, if $N(t)$ is infinite, there should exist an infinite cluster at time~0 in the constructed network.

Let $\lambda_{c}(t)$ denote the critical density corresponding to existence of an infinite cluster in the constructed network. If $t=0$, then the communication range of each node is $R$. Thus, $\lambda_{c}(t)=\lambda_{c}$ when $t=0$. As $t$ increases, the communication range of nodes in the constructed network, $2\cdot t\cdot v_{max}+R$, increases as well. Thus, $\lambda_{c}(t)$ is a decreasing function of $t$, and its global maximum occurs at $t=0$. $\lambda_{c}(t)$ approaches $\lambda_c$ as $t$ decreases. Thus, for each $\lambda <\lambda_{c}$, if we choose $t_0$ small enough, $\lambda_{c}(t_0)$ is greater than $\lambda$. If $\lambda_{c}(t_0)>\lambda$, there is no infinite cluster a.s. in the constructed network. Therefore, $N(t)$ is finite for $t\in(0,t_0]$ a.s. in the original network.
\end{IEEEproof}



\newtheorem{theorem}{Theorem}
\begin{theorem} \label{theorem:1}
In an infinite network, the probability that epidemic routing successfully delivers a message to the destination by time $T$, $T\ge 0$, experiences a phase transition at the percolation critical density, $\lambda_{c}$.
\end{theorem}

\begin{IEEEproof}
Label the infected nodes at time $t$, $i_1$, $i_2$, \dots, $i_{N(t)}$. First, we prove that if $N(t)$ is finite a.s. for some~$t>0$, the number of infected nodes at time~\mbox{$2t$}, $N(2t)$, is finite a.s. 
Let \mbox{$N^{j}(t,2t)$} denote the number of infected nodes at time~\mbox{$2t$} under the assumption that only $i_j$ and other nodes of the cluster to which $i_j$ belongs (if one exists) are infected at time~$t$. Since nodes are initially placed according to a homogeneous Poisson point process, and their mobility processes are i.i.d., node placement at any time~$t$ is a Poisson point process \cite{VANDENBERG1997247}. Thus, $N^{j}(t,2t)$ is equal to $N(t)$ in a second network that has the same set of nodes as the original network, and in which node~$i_j$ is the source. Moreover, initial positions of nodes in the second network are the same as at time~$t$ in the original network.
As a result, if $N(t)$ is finite a.s., $N^{j}(t,2t)$ is finite a.s. Node $i_j$ and all infected nodes to which the message has been forwarded along a path from $i_j$ in the interval~$(t,2t]$ are counted in \mbox{$N^{j}(t,2t)$}. Moreover, the message has been forwarded to each node infected by time~$2t$ along a path from a node infected by time~$t$. Thus, the number of infected nodes at time~$2t$, $N(2t)$, is at most $\sum_{j=1}^{N(t)} N^{j}(t,2t)$. If $N(t)$ is finite a.s., $\sum_{j=1}^{N(t)} N^{j}(t,2t)$ is finite a.s. since $N^{j}(t,2t)$, \mbox{$1\leq j\leq N(t)$}, is finite a.s. if $N(t)$ is finite a.s. as mentioned earlier. Therefore, $N(2t)$ is finite a.s. if $N(t)$ is finite a.s.

In a similar way, we can prove that if $N(t)$ and $N(l\cdot t)$ are finite a.s., $N((l+1)\cdot t)$ is finite a.s., $\forall l\in \mathbb{N}\setminus\{1\}$. By induction, we conclude that $N(l\cdot t)$ is finite a.s., $\forall l\in \mathbb{N}$, if $N(t)$ is finite a.s. Thus, if $N(t)$ is finite a.s., $N(\lceil\frac{T}{t}\rceil\cdot t)$ is finite a.s. Since $T\leq \lceil\frac{T}{t}\rceil\cdot t$, $N(T)$ is not greater than $N(\lceil\frac{T}{t}\rceil\cdot t)$. Therefore, $N(T)$ is finite a.s., $\forall~T>0$, if $N(t)$ is finite a.s. for some~$t$. According to Lemma~\ref{lemma:1}, for $\lambda <\lambda_{c}$, $\exists~t_0>0$ such that $N(t_0)$ is finite a.s. Thus, $N(T)$ is finite a.s., and $p(T)$ in an infinite network with density $\lambda <\lambda_{c}$ is zero. On the other hand, for each $\lambda>\lambda_{c}$, there exists a.s. one infinite cluster containing a constant fraction of nodes, $\forall t~0\leq t\leq T$ \cite{continuumPercolation, penrose2003random}. There is a positive probability that this infinite cluster which contains a constant fraction of nodes becomes infected in the interval $[0, T]$ (if it exists). Therefore, $p(T)$ in an infinite network with density \mbox{$\lambda > \lambda_{c}$} is positive, and a phase transition occurs at density~$\lambda_{c}$.
\end{IEEEproof}

\begin{figure}
\centering
\begin{tikzpicture}[scale=0.8]
    \begin{axis}[
    xlabel={Node Density, $\lambda$ (nodes/$km^{2}$)},
    ylabel={The Probability of Success, $p(T)$},
    xmin=300, xmax=650,
    ymin=0, ymax=1,
    xtick={300,350,400,450,500,550,600,650},
    ytick={0,0.2,0.4,0.6,0.8,1},
    legend pos=north west,
    ymajorgrids=true,
    grid style=dashed,
     every axis plot/.append style={thick}
    ]

    \addplot[
    color=brown,
    mark=diamond,
    mark options={solid},
    ]
    coordinates {
        (300,0.0019)
        (325,0.0024)
        (350,0.0031)
        (375,0.0041)
        (400,0.0062)
        (425,0.0095)
        (450,0.0162)
        (475,0.0298	)
        (500,0.0671)
        (525,0.1683)
        (550,0.4211) 
        (575,0.7161	)
        (600,0.8622)
        (625,0.9165)
        (650, 0.9487) 
    };\addlegendentry{$T=15~s$}
 
    \addplot[
    color=green,
    mark=triangle,
    mark options={solid},
    ]
    coordinates {
        (300,0.0024)
        (325,0.0031)
        (350,0.0041)
        (375,0.0056)
        (400,0.0090)
        (425,0.0140)
        (450,0.0250)
        (475,0.0490	)
        (500,0.1096)
        (525,0.3001)
        (550,0.6030) 
        (575,0.8272)
        (600,0.9217)
        (625,0.9522)
        (650,0.9696) 
    };\addlegendentry{$T=25~s$}
    
    \addplot[
    color=red,
    mark=circle,
    mark options={solid},
    ]
    coordinates {
        (300,0.0037)
        (325,0.0051)
        (350,0.0071)
        (375,0.0102)
        (400,0.0174)
        (425,0.0288)
        (450,0.0550)
        (475,0.1190)
        (500,0.2730)
        (525,0.6040)
        (550,0.8589) 
        (575,0.9462)
        (600,0.9724)
        (625,0.9853)
        (650,0.9901) 
    };\addlegendentry{$T=50~s$}
 
    \addplot[
    color=blue,
    mark=square,
    mark options={solid},
    ]
    coordinates {
        (300,0.0070)
        (325,0.0103)
        (350,0.0151)
        (375,0.0239)
        (400,0.0431)
        (425,0.0782)
        (450,0.1568)
        (475,0.3400	)
        (500,0.6455)
        (525,0.9028)
        (550,0.9761) 
        (575,0.9915)
        (600,0.9942)
        (625,0.9983)
        (650,0.9987) 
    };\addlegendentry{$T=100~s$}

    \addplot[
    color=black,
    mark=*,
    mark options={solid},
    ]
    coordinates {
        (300,0.0167)
        (325,0.0260)
        (350,0.0416)
        (375,0.0713)
        (400,0.1339)
        (425,0.2540)
        (450,0.4799)
        (475,0.7839)
        (500,0.9602)
        (525,0.9932)
        (550,0.9984) 
        (575,0.9996)
        (600,0.9998)
        (625,0.9999)
        (650,0.9999) 
    };\addlegendentry{$T=200~s$}

    \addplot[
    color=gray,
    mark=circle,
    mark options={solid},
    style=dashed,
    label=a,
    ]%
    coordinates {
        (572,0)
        (572,1)
    } node[pos=0.18, above right] {$\lambda=572$};
    
    \end{axis}
\end{tikzpicture}
\caption{Phase transition in the probability of success for different values of $T$ when $L=5$~\textit{km}, $R=50~m$, $v_{min}=0$, $v_{max}=1~m/s$, and $t_{r}=120~s$.} 
\label{fig:transitionInSp2}
\end{figure}
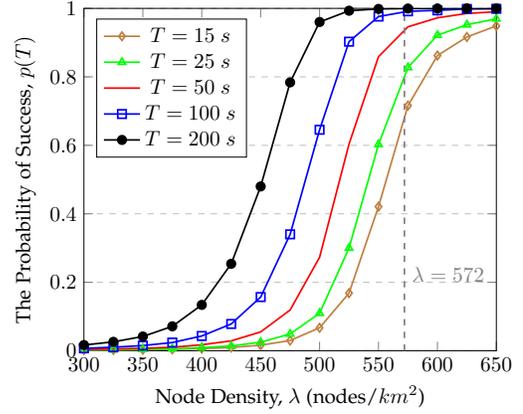

A large finite network exhibits a behavior similar to percolation \cite{haenggi2009stochastic}, and a very large cluster of nodes exists with a high probability in a supercritical large finite network. Thus, one can think that the transition of $p(T)$ in a finite network always occurs near the percolation critical density as in an infinite network. We present the probability of success for different deadlines, $T$, in a finite network in Fig.~\ref{fig:transitionInSp2}. According to (\ref{eq:threshold}), the percolation critical density for the setting applied to derive results presented in Fig.~\ref{fig:transitionInSp2} is greater than 572~nodes$/km^{2}$ with more than $99.99\%$ confidence. As it can be seen in this figure, for each value of $T$, the probability of success experiences a sudden change. As $T$ increases, the sudden change in the probability of success moves to the left. As a result, the sudden change corresponding to $T \geq 50$ in Fig.~\ref{fig:transitionInSp2} occurs completely in the subcritical regime. Thus, when the deadline is large enough, the probability of success experiences a sudden change in a subcritical finite network as density increases.

In Fig.~\ref{fig:transition}, as $L$ increases, the transition of $p(T)$ approaches line $\lambda=a$ where $a$ is in the interval $(565, 572)$ while the percolation critical density is in the interval $(572, 576)$ with more than $99.99\%$ confidence according to (\ref{eq:threshold}). Moreover, Fig.~\ref{fig:transitionInSp2} shows some phase transitions occurring in the subcritical regime.
Since the results presented in Figs.~\ref{fig:transition} and \ref{fig:transitionInSp2} are obtained from simulation of finite networks, the aforementioned observations are not in contradiction with Theorem~\ref{theorem:1} that implies that in an infinite network, the phase transition of $p(T)$ occurs at the percolation critical density. As it can be seen in Fig.~\ref{fig:transitionInSp2}, the deadline affects the range of densities at which $p(T)$ undergoes a phase transition in a finite network. For example, if we choose value of $T$ sufficiently small to derive results presented in Fig.~\ref{fig:transition}, the curves for different values of $L$ intersect in the interval $(572, 575)$.
However, the transition of $p(T)$ in an infinite network occurs at the percolation critical density regardless of the value of deadline. According to Theorem~\ref{theorem:1}, an infinite number of nodes cannot be infected within a finite deadline in an subcritical infinite network a.s. Thus, in contrast to a finite network, no phase transition occurs in the subcritical regime when the network is infinite.

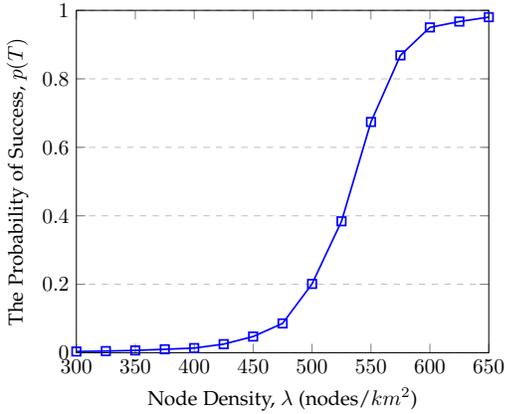
\begin{figure}
\centering
\begin{tikzpicture}[scale=0.8]
\begin{axis}[
    xlabel={Node Density, $\lambda$ (nodes/$km^{2}$)},
    ylabel={The Probability of Success, $p(T)$},
    xmin=300, xmax=650,
    ymin=0, ymax=1,
   xtick={300,350,400, 450, 500, 550, 600, 650},
    ytick={0,0.2,0.4,0.6,0.8,1},
    legend pos=north west,
    ymajorgrids=true,
    grid style=dashed,
     every axis plot/.append style={thick}
]

      \addplot[
    color=blue,
    mark=square,
    mark options={solid},
    ]
   coordinates {
        (300,27.14952380952381/7499)
        (325,39.446216216216214/8124)
        (350,59.10645161290323/8749)
        (375, 94.87481481481481/9374)
        (400,136.48/9999)
        (425,267.41/10624)
        (450,532.5/11249)
        (475,1016.86/11874)
        (500,2513.39/12499)
        (525,5039.785/13124)
        (550,9267.833333333334/13749)
        (575,12484.086/14374)
        (600,14253.405714285715/14999)
        (625,15114.73/15624)
        (650,15928.298571428571/16249)
    };

\end{axis}
\end{tikzpicture}
\caption{The probability of success for $T=50~s$, $L=5~km$, $R=50~m$, $v_{min}=0$, $v_{max}=1~m/s$, $t_{r}=120~s$, and $d=0.1~s$.
} 
\label{fig:transition2}
\end{figure}

As mentioned earlier, simulation results presented in Fig.~\ref{fig:transitionInSp2} show that $p(T)$ exhibits a phase transition when transfer delays are zero. Similar simulation results show that $p(T)$ exhibits a phase transition when transfer delays are positive. For example, assume that transfer delay, denoted by $d$, is 0.1 \textit{s}. Fig.~\ref{fig:transition2} represents PS of epidemic routing when $T=50~s$, $L=5~km$, $R=50~m$, $v_{min}=0$, $v_{max}=1~m/s$, $t_{r}=120~s$, and $d=0.1~s$. As observed in this figure, PS, $p(T)$, exhibits a phase transition as the density of nodes increases even when transfer delays are positive. 

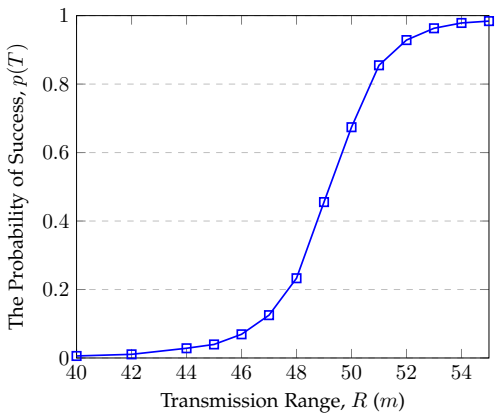
\begin{figure}
\centering
\begin{tikzpicture}[scale=0.8]
\begin{axis}[
    xlabel={Transmission Range, $R$ ($m$)},
    ylabel={The Probability of Success, $p(T)$},
    xmin=40, xmax=55,
    ymin=0, ymax=1,
   xtick={40, 42, 44, 46, 48, 50, 52, 54},
    ytick={0,0.2,0.4,0.6,0.8,1},
    legend pos=north west,
    ymajorgrids=true,
    grid style=dashed,
     every axis plot/.append style={thick}
]

      \addplot[
    color=blue,
    mark=square,
    mark options={solid},
    ]
   coordinates {
        (40, 80.40410256410256/13749)
        (42,146.125/13749)
        (44, 389.89/13749)
        (45, 543.36/13749)
        (46,951.035/13749)
        (47, 1719.564375/13749)
        (48,3202.39/13749)
        (49, 6261.347142857143/13749)
        (50, 9267.833333333334/13749)
        (51,11754.29/13749)
        (52,12767.11/13749)
        (53,13240.2415/13749	)
       (54,	13453.704210526315/13749)
        (55,13529.361481481481/13749)
    };

\end{axis}
\end{tikzpicture}
\caption{The probability of success for $T=50~s$, $L=5~km$, $\lambda=550$~nodes/$km^{2}$, $v_{min}=0$, $v_{max}=1~m/s$, $t_{r}=120~s$, and $d=0.1~s$.
} 
\label{fig:transitionUpanChangingR}
\end{figure}

If $\lambda$ is fixed, a phase transition associated with the existence of an infinite cluster occurs as $R$ increases. If $R$ is less (more) than the critical transmission range in an infinite network, the network is subcritical (supercritical), and $p(T)$ is zero (positive) according to the proof of Theorem~\ref{theorem:1}. Therefore, the probability of success under epidemic routing in an infinite network exhibits a phase transition also as $R$ increases. Similarly, a phase transition can be observed in the probability of success under epidemic routing in a finite network upon increasing the transmission range. For example, Fig.~\ref{fig:transitionUpanChangingR} presents the results of the probability of success when $T=50~s$, $L=5~km$, $\lambda=550$~nodes/$km^{2}$, $v_{min}=0$, $v_{max}=1~m/s$, $t_{r}=120~s$, $d=0.1~s$, and $R$ varies from 40~\textit{m} to 55~\textit{m}. As observed in this figure, the probability of success exhibits a phase transition as $R$ increases. As mentioned in Section~\ref{sec:intro}, phase transition phenomena can be exploited to improve the efficiency of networks. Although it is hard to control node density, the transmission range can be controlled through changing signal strength. 


\subsection{Discussion}
It is interesting to observe how the probability of success behaves under a more realistic mobility model. In this section, we show that the probability of success exhibits a phase transition upon increasing node density when nodes move according to the TVC mobility model proposed in \cite{4215676}. Some important characteristics, observed in the wireless LAN traces, such as skewed location visiting preferences are captured by this model. According to this model, nodes move in an $L\times L$ simulation area. Each community is an $L_c\times L_c$ square in the simulation area and is frequently visited by a node. Communities are chosen with uniform distribution in the simulation area.

\begin{figure}
    \centering
    \resizebox{0.54\columnwidth}{!}{
\includegraphics[]{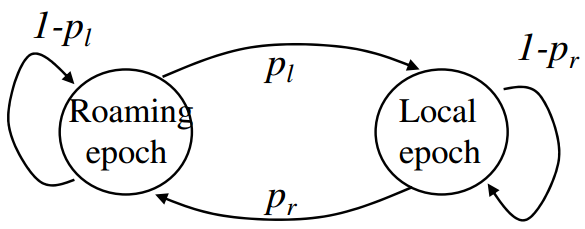}}
    \caption{The Markov chain governing choosing of the movement modes in the TVC mobility model.}
    \label{fig:epochs}
\end{figure}

\noindent
There are two different movement modes for nodes, called \textit{local epoch} and \textit{roaming epoch}, that can be considered a travel of the random direction mobility model in a community and simulation area, respectively. The durations of the local and roaming epochs are exponentially distributed with averages of $\overline{D}_l$ and $\overline{D}_r$, respectively.
When each epoch ends, the node pauses for a duration which is chosen uniformly from $[0, P_{max}]$. Afterwards, the node chooses the next movement node according to the Markov chain presented in Fig.~\ref{fig:epochs}. Let consider a node that is outside the community which is assigned to it and has just chosen the local epoch as its next movement mode.
In order to preserve continuity of movement of such a node, a transitional epoch is considered in the TVC mobility model. Such a node chooses a random position with uniform distribution in the community that is assigned to this node, and move along the shortest straight path to reach the chosen position during transitional epoch.

\begin{figure}
\centering
\begin{tikzpicture}[scale=0.8]
    \begin{axis}[
        xlabel={Node Density, $\lambda$ (nodes/$m^{2}$)},
        ylabel={The Probability of Success, $p(T)$},
        xmin=0.001, xmax=0.011,
        ymin=0, ymax=1,
        xtick={0.001, 0.002, 0.003, 0.004,0.005,0.006 ,0.007 ,0.008, 0.009 ,0.01 ,0.011},
        ytick={0,0.2,0.4,0.6,0.8,1},
        legend pos=north west,
        ymajorgrids=true,
        grid style=dashed,
         every axis plot/.append style={thick}
    ]

      \addplot[
    color=blue,
    mark=square,
    mark options={solid},
    ]
   coordinates {
        (0.001, 0.010752858)
        (0.002, 0.024812406)
        (0.003, 0.039735781)
        (0.004,	0.054563641)
        (0.005, 0.117493499)
        (0.006, 0.218809385)
        (0.007, 0.492505179)
        (0.008, 0.834206776)
        (0.009,	0.952723636)
        (0.010, 0.987479548)
        (0.011, 0.997753782)
    };\addlegendentry{$T=100~s$}

	  \addplot[
    color=red,
    mark=triangle,
    mark options={solid},
    ]
   coordinates {
        (0.001, 0.003034614)
        (0.002, 0.005138283)
        (0.003, 0.007769256)
        (0.004,	0.012603151)
        (0.005, 0.023369674)
        (0.006, 0.039744957)
        (0.007, 0.084452243)
        (0.008, 0.212272367)
        (0.009,	0.485808053)
        (0.010, 0.857743989)
        (0.011, 0.978087285)
    };\addlegendentry{$T=50~s$}
\end{axis}
\end{tikzpicture}
\caption{The probability of success when deadline is 50~\textit{s} or 100~\textit{s}, and nodes move according to the TVC model.} 
\label{fig:transitionMoreRealisticModel}
\end{figure}
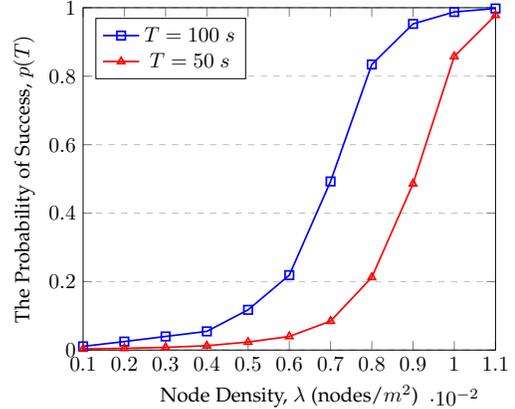

\noindent
Fig.~\ref{fig:transitionMoreRealisticModel} represents the results of the probability of success obtained from the simulation when nodes move according to the TVC model with parameters $L=1000$~\textit{m}, \mbox{$L_c=100$~\textit{m}}, $R=10$~\textit{m}, $v_{min}=1$~\textit{m/s}, $v_{max}=3$~\textit{m/s}, $\overline{D}_l=100$~\textit{s}, $\overline{D}_r=500$~\textit{s}, $P_{max}=100$~\textit{s}, $p_l=0.8$, and $p_r=0.2$. As observed in Fig.~\ref{fig:transitionMoreRealisticModel}, the probability of success exhibits a phase transition as node density increases.


\section{Proposed ODE Models}\label{sec:pairwiseRate}
Previous ODE models of epidemic routing are based on the assumption that mobile nodes do not form clusters of size greater than two~\cite{zhang2007performance, banerjee2008relays, spyropoulos2009routing}, and have been shown to be accurate when networks are sparse. Although this assumption is reasonable for sparse networks, it is not reasonable for non-sparse networks where large clusters can form with non-negligible probability. In this section, we first introduce the standard ODE model proposed in \cite{zhang2007performance} for epidemic routing in sparse networks. We show that it is inaccurate when used to model non-sparse networks. We then propose four new ODE models.

The following ODE model was proposed in \cite{zhang2007performance} for basic epidemic routing,
\begin{equation}\label{eq:ode}
\frac{d\overline{N}(t)}{dt}=\beta\cdot \overline{N}(t)\cdot(M - \overline{N}(t)),
\end{equation} where $\beta$, $\overline{N}(t)$, and $M$ are the pairwise meeting rate, the average number of infected nodes at time $t$, and the total number of nodes in the network, respectively. (\ref{eq:ode}) has the following solution,
\begin{equation}\label{eq:N(t)}
\overline{N}(t)=\frac{M\cdot \overline{N}(0)}{(M-\overline{N}(0))\cdot e^{-\beta\cdot M\cdot t}+\overline{N}(0)},
\end{equation} where $\overline{N}(0)$ is the number of initial infected nodes, which is one in a network with single source. 
Recall that we are interested in $p(T)$, the probability that the message is received by the destination by time~$T$. It can be computed using (\ref{eq:N(t)}) as follows,
\begin{equation}\label{eq:sp}
\centering
\begin{split}
p(T)&=\frac{\overline{N}(T)-1}{M-1}\\
&=\frac{(M-1)\cdot \overline{N}(0)-(M-\overline{N}(0))\cdot e^{-\beta\cdot M\cdot T}}{(M-1)\cdot((M-\overline{N}(0))\cdot e^{-\beta\cdot M\cdot T}+\overline{N}(0))}.
\end{split}
\end{equation}

The model given in (\ref{eq:ode}) focuses on modeling meetings between one infected node and one uninfected node. When such a meeting occurs, the uninfected node becomes infected after a short delay, usually assumed to be zero in the literature \cite{zhang2007performance, picu2015dtn, wang2015restricted}. Thus, the number of infected nodes increases by one.
This modeling approach does not account for the possibility of an infected node meets a large cluster as can occur in a non-sparse network. When an infected node meets a node in an uninfected large cluster, the message is distributed to all nodes of the cluster quickly, and more than one node is infected in a small time interval. Moreover, Eq.~(\ref{eq:ode}) is valid if the meeting events between one infected node and one uninfected node are independent of each other, and only one node becomes infected during each meeting event. However, neither condition holds in a non-sparse network. In the following, we present some results, obtained from Eq.~(\ref{eq:ode}) and simulation, that indicate the model given in (\ref{eq:ode}) cannot be applied to evaluate the performance of epidemic routing in non-sparse networks.  

\begin{figure}
\centering
\begin{tikzpicture}[scale=0.8]
    \begin{axis}[
    xlabel={The Deadline, $T$ ($s$)},
    ylabel={The Probability of Success, $p(T)$},
    xmin=0, xmax=120,
    ymin=0, ymax=1,
    xtick={0,20,40,60,80,100,120},
    ytick={0,0.2,0.4,0.6,0.8,1},
    legend style={at={(0.98,0.85)}},
    ymajorgrids=true,
    grid style=dashed,
     every axis plot/.append style={very thick}
    ]
    
      \addplot[
    color=blue,
    mark=circle,
    mark options={solid},
    ] table [x=time, y=simulation, col sep=space] {comparison.txt};\addlegendentry{\normalsize Simul.\textsuperscript{0}}
    
    \addplot[
    color=green,
    mark=circle,
    mark options={solid},
    ] table [x=time, y=simulPosDelay, col sep=space] {comparison.txt};\addlegendentry{\normalsize Simul.\textsuperscript{+}}

    \addplot[
  color=black,
   mark=circle,
    mark options={solid},
    style=loosely dotted
   ]
 table [x=time, y=meetingRate, col sep=space] {comparison.txt};\addlegendentry{\normalsize $\beta=~$\small PMR}

    \addplot[
  color=brown,
    mark=circle,
    mark options={solid},
    style=dashed
    ]
 table [x=time, y=optLambdaZeroDelay, col sep=space] {comparison.txt};\addlegendentry{\normalsize $\beta=\beta_{opt}$}
 
    \addplot[
    color=red,
   mark=circle,
    mark options={solid},
    style=densely dotted
    ]
 table [x=time, y=averageBeta, col sep=space] {comparison.txt};\addlegendentry{\normalsize $\beta=\overline{\beta}$}

    \end{axis}
\end{tikzpicture}
\caption{Results of the probability of success obtained from the simulation and computed by~Eq.~(\ref{eq:sp}) for different values of~$\beta$. Curves \textit{Simul.\textsuperscript{0}} and \textit{Simul.\textsuperscript{+}} correspond to the cases of 0 delays and non-zero delays, respectively.} 
\label{fig:poorFit}
\end{figure}
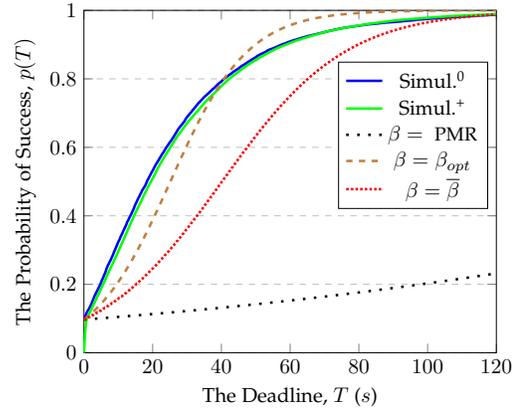

As mentioned earlier, in a non-sparse network, large clusters can significantly accelerate message dissemination.
Moreover, the source node may belong to a non-negligible size cluster at time 0 in which case all nodes in the cluster are infected at time 0 when transfer delays are negligible. Thus, $\overline{N}(0)$ in a non-sparse network is greater than one in the case of zero transfer delay. The ODE given in (\ref{eq:ode}) does not work well for non-sparse networks. For example, Fig.~\ref{fig:poorFit} presents the simulation results for $p(T)$ and results computed by Eq.~(\ref{eq:sp}) using a pairwise meeting rate~$\beta$ and $\overline{N}(0)$ taken from a simulation of a network with parameters $\lambda=550$~nodes/$km^{2}$, $L=5~km$, $R=50~m$, $v_{min}=0$, \mbox{$v_{max}=1~m/s$}, and $t_{r}=120~s$. Results are presented both for the case of 0 delays and non-zero delays, where nodes use WiFi-direct with transfer rate 50~\textit{Mbps} for each connection, and message sizes of 25 \textit{KB}.
In the case of non-zero delays, although the probability of success obtained from the simulation is 0 when $T=0$, it increases fast as $T$ increases, and approaches that for 0 transfer delay.
As observed in Fig.~\ref{fig:poorFit}, when $\beta$ is set to the pairwise meeting rate (PMR), the results computed by Eq.~(\ref{eq:sp}) do not match simulation results well. However, motivated by this ODE model, we ask if it can be adapted to a non-sparse network.

One can consider other ways of setting parameter~$\beta$ in~(\ref{eq:ode}). 
In order to achieve high accuracy, according to (\ref{eq:ode}) $\beta$ should be a good approximation of function~\mbox{$\frac{d\overline{N}(t)}{dt}/(\overline{N}(t)\cdot(M-\overline{N}(t)))$}. One option is to use $\overline{\beta}$, defined below, in order to model the network in the interval~$[0,T_{max}]$.
\begin{equation}\label{eq:averageBeta}
  \overline{\beta}=\frac{1}{T_{max}}\int_{0}^{T_{max}} \frac{\overline{N}'(t)}{\overline{N}(t)\cdot(M-\overline{N}(t))} dt,
\end{equation} where $\overline{N}'(t)$ ($d\overline{N}(t)/dt$) and $\overline{N}(t)$ are obtained from simulation. For example, simulation results for the previously mentioned network suggest $\overline{\beta}=4.04\times 10^{-6}$ for \mbox{$T_{max}=120~s$}. The results computed by (\ref{eq:sp}) for $\beta=\overline{\beta}$ are presented in Fig.~\ref{fig:poorFit}. As observed in this figure, using $\overline{\beta}$ in the ODE model (in Eq.~(\ref{eq:ode})) produces a significant error.

One can ask if \textit{any} constant $\beta$ exists such that the ODE model given in (\ref{eq:ode}) is accurate for non-sparse networks. In order to answer to this question,
we determine the value of~$\beta$ that minimizes the Integrated Squared Error of the results computed from (\ref{eq:sp}) with respect to results obtained from simulation, denoted by $\beta_{opt}$. For example, $\beta_{opt}=6.46\times 10^{-6}~s^{-1}$ for the previously mentioned network and $T\in[0,120]$. Analytic results computed using $\beta_{opt}$ in (\ref{eq:sp}) are presented in Fig.~\ref{fig:poorFit}. As observed in Fig.~\ref{fig:poorFit}, even $\beta_{opt}$ which minimizes error leads to a poor fit. Therefore, there exists no constant~$\beta$ for which the ODE model given in (\ref{eq:ode}) works for the aforementioned network.

In order to propose a performance model for epidemic routing in non-sparse networks, we introduce the notion of \textit{pairwise infection rate}, defined as the infection rate per pair of nodes that includes one infected node and one uninfected node. It is computed by dividing the rate at which uninfected nodes become infected, termed \textit{infection rate}, $\frac{d\overline{N}(t)}{dt}$, by the total number of pairs in the network that include one infected node and one uninfected node.

The infection rate depends on many factors including the number of infected nodes, time, shapes of clusters, relative positions of infected and uninfected clusters, and mobility parameters. The aforementioned parameters, network dynamics, and large number of nodes make deriving an exact formula for the pairwise infection rate too complex. Thus, we aim to approximate the infection rate and the pairwise infection rate. One alternative is to approximate them as a function of the number of infected nodes as defined below,
\begin{equation}\label{eq:betaN}
\beta_{num}(N)=\frac{R_{num}(N)}{N\cdot(M-N)},
\end{equation} where $R_{num}(N)$ is the infection rate as a function of the number of infected nodes. 
Another alternative is to approximate the infection rate and the pairwise infection rate as a function of time as given below,
\begin{equation}\label{eq:betat}
\beta_{time}(t)=
\frac{R_{time}(t)}{\overline{N}(t)\cdot(M-\overline{N}(t))},
\end{equation} where $R_{time}(t)$ is the infection rate at time $t$ as a function of time.

%

In order to characterize $\beta_{num}$ and $\beta_{time}$, we conducted a set of discrete-event simulations of non-sparse networks. In these simulations, we choose the time step, $\Delta t$, small enough to capture individual meetings of nodes, and transfer delays are neglected. Subsequently, we used the curve fitting toolbox of Matlab to obtain $\beta_{num}$ and $\beta_{time}$. In the rest of this section, we present a detailed description of the simulations and our characterization of $\beta_{num}$ and $\beta_{time}$. We then introduce new ODE models based on the obtained results. 

\subsection{Characterization of $\boldsymbol{\beta_{num}}$}\label{sec:betaN}
We simulate the network from time 0 until the time at which all nodes of the network are infected. We are especially interested in behavior of $\beta_{num}$ from \mbox{$N=\lfloor\overline{N}(0)\rfloor$} to $N=M-1$. As mentioned earlier, $\overline{N}(0)$ is the average number of infected nodes at time 0, and is obtained from simulation of non-sparse networks since the source initially infects the whole cluster to which it belongs. Let $r_{1}$, $r_{2}$, \dots, $r_{s}$ denote the simulation runs where $s$ is the number of runs. In simulation run $r_{i}$, \mbox{$1\leq i\leq s$}, for each~\mbox{$j\in\{\lfloor\overline{N}(0)\rfloor, \dots, M-1\}$}, we record the first and last times at which the observed number of infected nodes is $j$, denoted by $t^{f}_{i,j}$ and $t^{l}_{i,j}$, respectively. Moreover, the number of infected nodes at time $t^{l}_{i,j}+\Delta t$ is recorded and is denoted by~$N_{i,j}$. There are $(t^{l}_{i,j}-t^{f}_{i,j})/\Delta t$~time steps between $t^{f}_{i,j}$ and $t^{l}_{i,j}$. The change in the number of infected nodes during each of these steps is zero whereas the number of infected nodes increases by $N_{i,j}-j$ during the time step beginning at $t^{l}_{i,j}$. Let $\Delta\overline{N}_{N=j}$, $\lfloor\overline{N}(0)\rfloor\leq j\leq M-1$, denote the average change in the number of infected nodes during the next time step when there are currently $j$ infected nodes in the network. According to the explanation given above, $\Delta\overline{N}_{N=j}$, can be computed as follows,
\begin{equation} \label{eq:deltaN}
\centering
\Delta\overline{N}_{N=j}= \frac{\sum_{i=1}^{s} (N_{i,j}-j)}{\sum_{i=1}^{s}(\frac{t^{l}_{i,j}-t^{f}_{i,j}}{\Delta t}+1)}.
\end{equation}
Finally, using $\Delta\overline{N}_{N=j}$ and (\ref{eq:betaN}), $R_{num}$ and $\beta_{num}$ are estimated as
\begin{equation}
\centering
R_{num}(j)~\hat{=}~\frac{\Delta \overline{N}_{N=j}}{\Delta t}.
\end{equation}
\begin{equation} \label{eq:LambdaN}
\centering
\beta_{num}(j)~\hat{=}~\frac{\Delta \overline{N}_{N=j}}{\Delta t\cdot j\cdot(M-j)}.
\end{equation}

\begin{figure}
\centering
\begin{subfigure}{1\columnwidth}
\resizebox{1\columnwidth}{!}{
\includegraphics[]{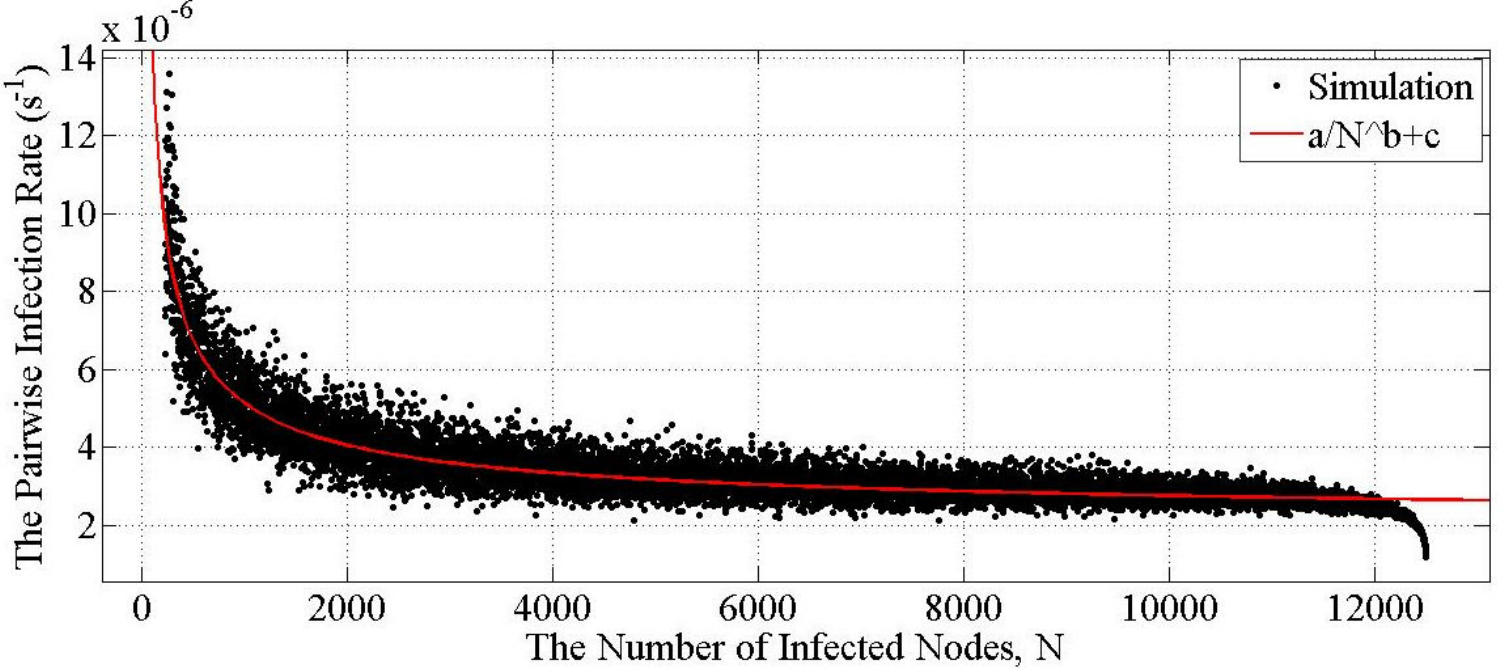}}
\caption{$\lambda=500$~nodes/$km^{2}$}
\end{subfigure}

\begin{subfigure}{1\columnwidth}
\resizebox{1\columnwidth}{!}{
\includegraphics[]{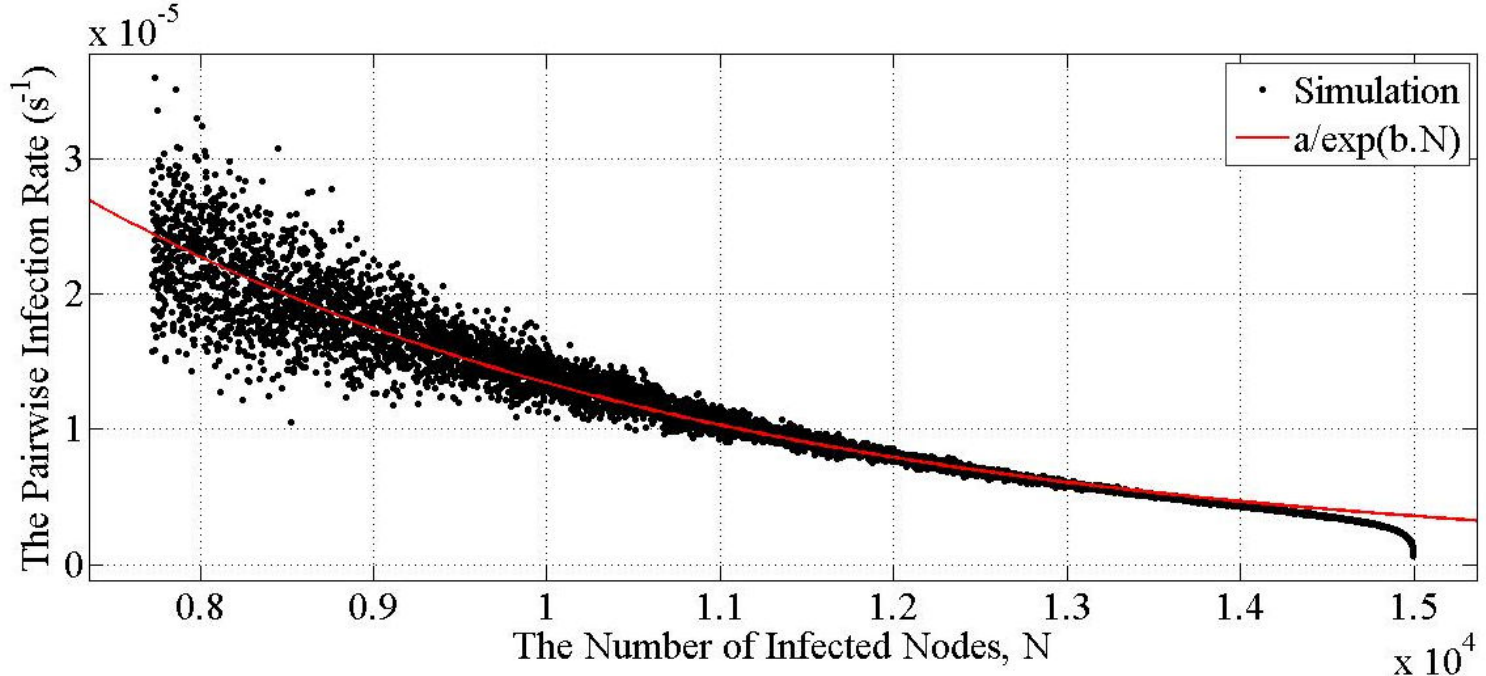}}
\caption{$\lambda=600$~nodes/$km^{2}$}
\end{subfigure}
\caption{Fitting curve to pairwise infection rate $\beta_{num}$ which is obtained from simulation where $L=5~km$, $R=50~m$, $v_{min}=0$, $v_{max}=1~m/s$, and $t_{r}=120~s$.}
\label{fig:LambdaN}
\end{figure}

The results obtained from simulation for $\beta_{num}$ show that $\beta_{num}$ exhibits a power law behavior and an exponential behavior in subcritical and supercritical networks, respectively. For example, Figs.~\ref{fig:LambdaN}(a) and \ref{fig:LambdaN}(b) present results for $\beta_{num}$ for two networks with densities \mbox{$\lambda=500$~nodes/$km^{2}$} and \mbox{$\lambda=600$~nodes/$km^{2}$}, which are considered subcritical and supercritical, respectively, since the critical density, $\lambda_{c}$, is 572~nodes/$km^{2}$ for \mbox{$R=50~m$} \cite{quintanilla2000efficient}. We note that Fig.~\ref{fig:LambdaN}(b) represents simulation results beginning from $N=\lfloor\overline{N}(0)\rfloor=7722$, which is large because the corresponding network is supercritical, and a large fraction of nodes belong to one cluster with high probability~\cite{haenggi2009stochastic}. If the source belongs to this cluster, a large fraction of nodes become infected at time 0. As observed in Fig.~\ref{fig:LambdaN}(a), the simulation results for $\beta_{num}$, except when $N$ is close to the total number of nodes, interval $(12000, 12500)$, exhibits a power law behavior, and the curve $a\cdot N^{-b}+c$ fits simulation results. Moreover, as observed in Fig. \ref{fig:LambdaN}(b), $\beta_{num}$ exhibits an exponential behavior excluding interval $(14000,15000)$, and the curve $a\cdot e^{-b\cdot N}$ fits simulation results. Table~\ref{tab:fittingLambdaN} represents the values of parameters and error for fitted curves in Fig.~\ref{fig:LambdaN}.

\begin{table}
    \renewcommand{\arraystretch}{1.5}
	\setlength{\tabcolsep}{3pt}
	\begin{center}
	\caption{The values of parameters and the root mean square error (RMSE) for the fitted curves in Fig.~\ref{fig:LambdaN}.}
	\label{tab:fittingLambdaN}
	\end{center}
	\centering
	\begin{tabular}{c|c|c|c|c}
	
		\hline
		$\boldsymbol{\lambda}$~(nodes/$km^{2}$) & $\boldsymbol{a}$ & $\boldsymbol{b}$ & $\boldsymbol{c}$ & \bfseries RMSE\\

	     \hline
	     \hline
	    \bfseries 500 &  0.000201 &   0.6009 &  1.971e-6 &  4.414e-7\\
	     \hline
	     \bfseries 600 &   0.0001875 &  0.0002639 & - & 1.543e-6\\
	     \hline
	     
	\end{tabular}
\end{table}

Thus, we approximate $\beta_{num}(N)$ respectively for subcritical and supercritical networks as
\begin{equation} \label{eq:estimatedLambdaN_sub}
\centering
\beta_{num}(N)~\hat{=}~a\cdot N^{-b}+c,
\end{equation}
\begin{equation} \label{eq:estimatedLambdaN_super}
\centering
\beta_{num}(N)~\hat{=}~a\cdot e^{-b\cdot N},
\end{equation}
where $a$, $b$, and $c$ are obtained from simulation. We have also tried an exponential fit, namely $a'\cdot e^{-b'\cdot N}+c'$ for subcritical networks and a power law fit, namely $a'\cdot N^{-b'}+c'$ for supercritical networks, but even using the best values for $a'$, $b'$, and $c'$, they produce larger errors than the power law and exponential fits, respectively. Moreover, We have tried to fit an appropriate curve to $R_{num}$, but the resulting error was significantly more than the case of $\beta_{num}$.

\subsection{Characterization of $\boldsymbol{R_{time}}$ and $\boldsymbol{\beta_{time}}$}
We simulate the network from time 0 until the time at which almost all nodes of the network (around 99 percent of nodes) are infected, denoted by $T_l$. In each simulation run~$r_i$, $1\leq i\leq s$, for each time $t_{j}=j\cdot\Delta t$, $j\in\{$0, 1, \dots, $\lfloor\frac{T_l}{\Delta t}\rfloor\}$, the total number of nodes infected by time~$t_{j}$, denoted by $N_{i}(t_{j})$, is logged. 
Afterwards, we compute the average total number of infected nodes at time $t_{j}$ as \mbox{$\overline{N}_{t=t_{j}}=\frac{\sum_{i=1}^{s}N_{i}(t_{j})}{s}$}. The average number of nodes that become infected during the interval $(t_{j},t_{j+1}]$, denoted by $\Delta\overline{N}_{t=t_{j}}$, equals \mbox{$\overline{N}_{t=t_{j+1}}-\overline{N}_{t=t_{j}}$}, \mbox{$0\leq j< \lfloor\frac{T_l}{\Delta t}\rfloor$}. Thus, $R_{time}$ is estimated as
\begin{equation}\label{eq:R_time}
\centering
R_{time}(t_{j})~\hat{=}~\frac{\Delta \overline{N}_{t=t_{j}}}{\Delta t}.
\end{equation} According to Eqs.~(\ref{eq:betat}) and (\ref{eq:R_time}) and using $\overline{N}_{t=t_{j}}$, $\beta_{time}$ is estimated as
\begin{equation} \label{eq:Lambdat}
\centering
\beta_{time}(t_{j})~\hat{=}~\frac{\Delta \overline{N}_{t=t_{j}}}{\Delta t\cdot \overline{N}_{t=t_{j}}\cdot(M-\overline{N}_{t=t_{j}})}.
\end{equation}

\begin{figure}
\centering
\begin{subfigure}{1\columnwidth}
\resizebox{1\columnwidth}{!}{
\includegraphics[]{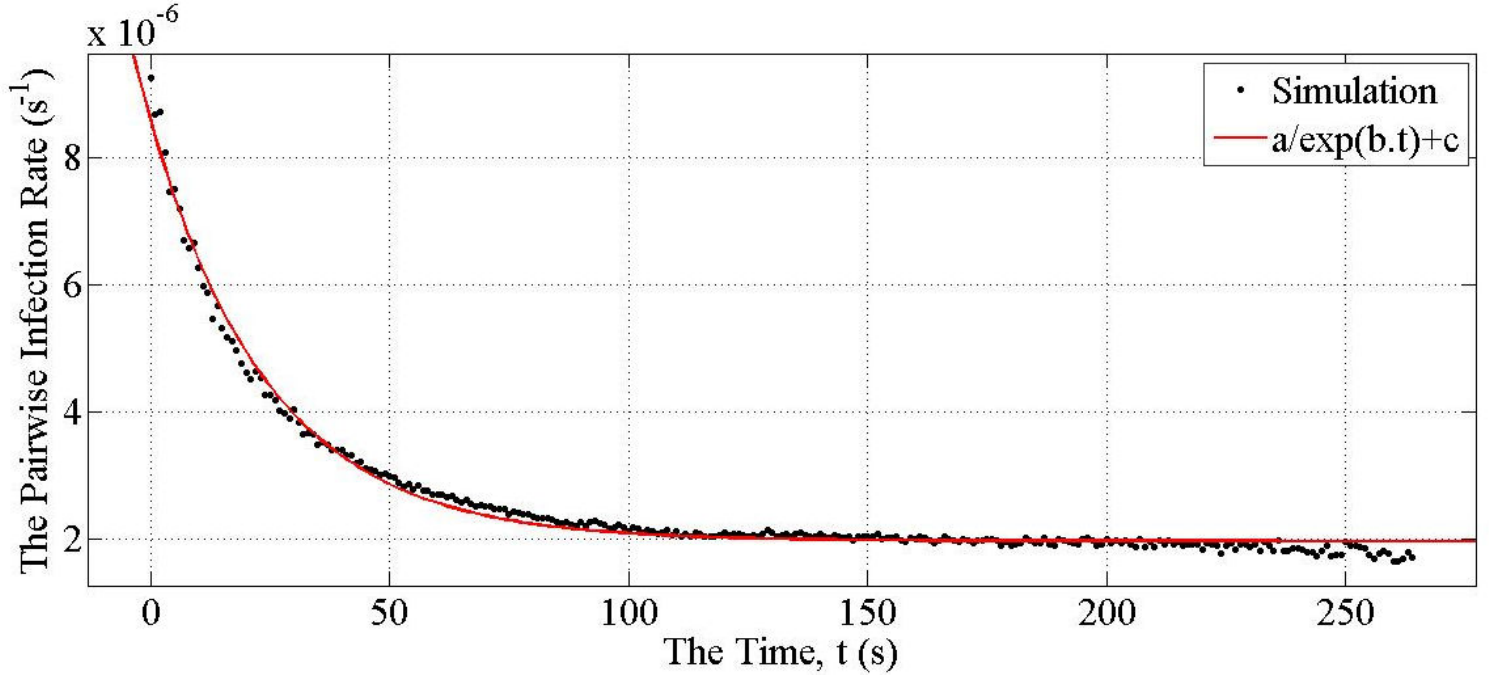}}
\caption{$\lambda=500$~nodes/$km^{2}$}
\end{subfigure}
\begin{subfigure}{1\columnwidth}
\resizebox{1\columnwidth}{!}{
\includegraphics[]{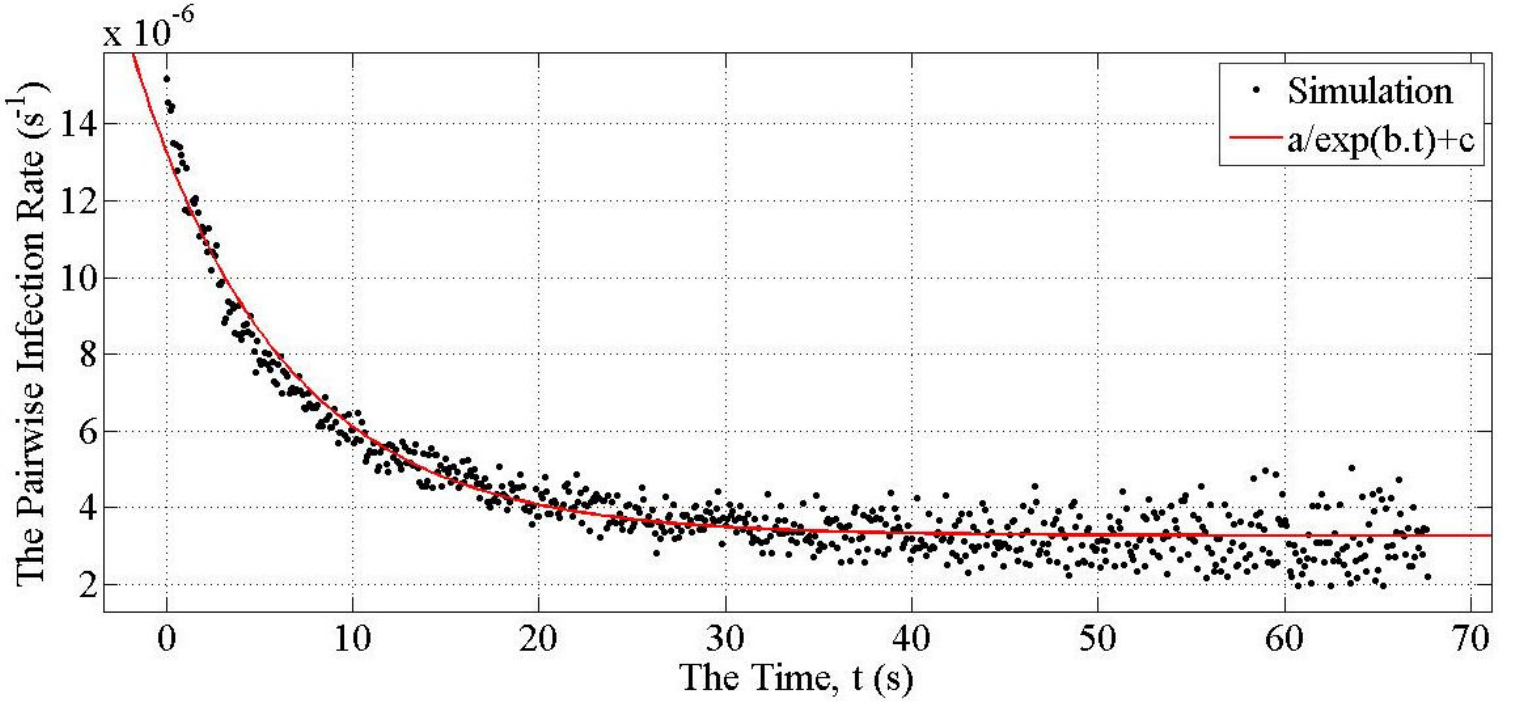}}
\caption{$\lambda=600$~nodes/$km^{2}$}
\end{subfigure}
\caption{Fitting curve to pairwise infection rate $\beta_{time}$ which is obtained from simulation where $L=5~km$, $R=50~m$, $v_{min}=0$, $v_{max}=1~m/s$, and $t_{r}=120~s$.}
\label{fig:Lambdat}
\end{figure}

The results obtained from simulation for $\beta_{time}$ show that it decays exponentially. 
For example, Fig.~\ref{fig:Lambdat} represents the results for $\beta_{time}$ in the networks with densities \mbox{$\lambda=500~\text{nodes/}km^{2}$} and $\lambda=600$~nodes/$km^{2}$. As observed in Fig.~\ref{fig:Lambdat}, simulation results for $\beta_{time}$ exhibit an exponential decay, and the curve $a\cdot e^{-b\cdot t}+c$ fits the results well. Thus, function $\beta_{time}(t)$ can be estimated as
\begin{equation} \label{eq:estimatedLambdat}
\centering
\beta_{time}(t)~\hat{=}~a\cdot e^{-b\cdot t}+c,
\end{equation} where parameters $a$, $b$, and $c$ are obtained from simulation. It is worth mentioning that an exponential curve fits $\beta_{time}$ both when network is subcritical and supercritical unlike $\beta_{num}$.
We have also tried fitting a power law expression, namely \mbox{$a'\cdot t^{-b'}+c'$}, excluding $t=0$, but it produces larger errors. 

\begin{figure}
\centering
\resizebox{1\columnwidth}{!}{
\includegraphics[]{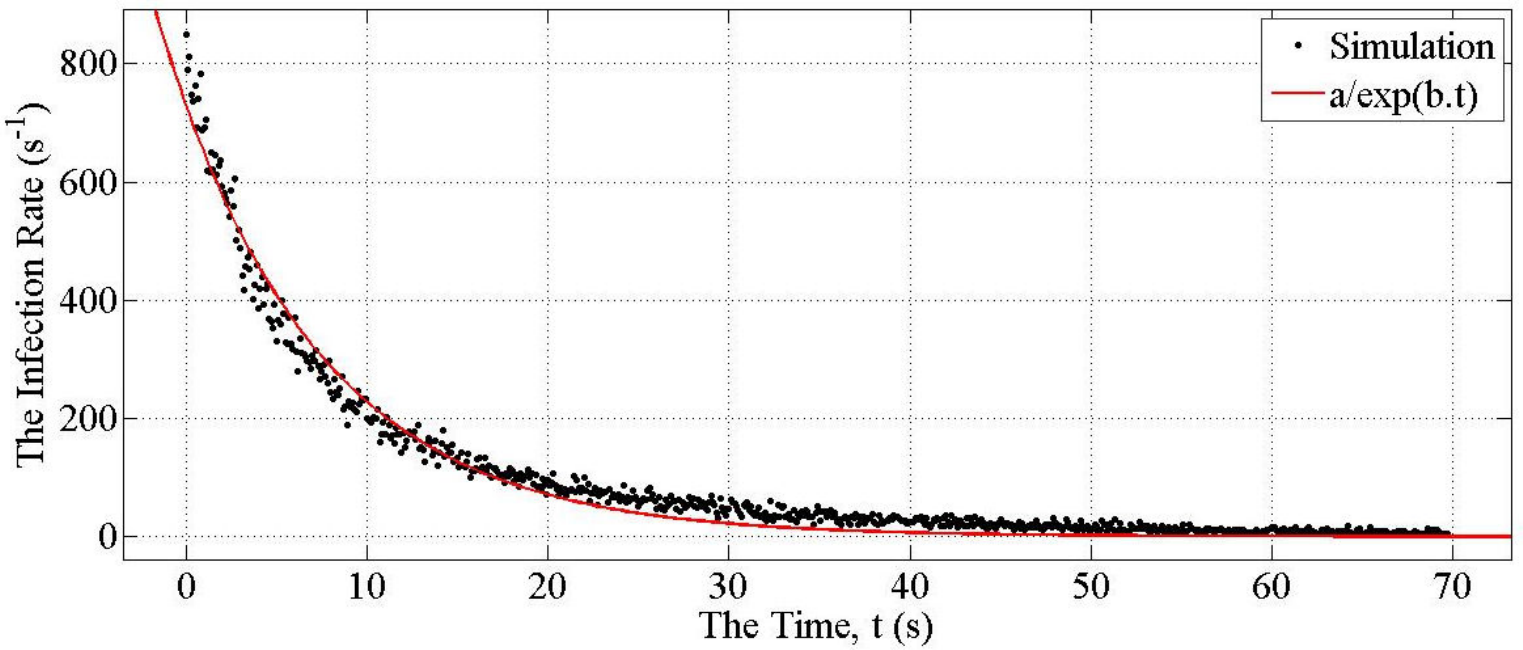}}
\caption{Fitting curve to infection rate $R_{time}$ which is obtained from simulation where $\lambda=600$~nodes/$km^2$, $L=5~km$, $R=50~m$, $v_{min}=0$, $v_{max}=1~m/s$, and $t_{r}=120~s$.}
\label{fig:R_time}
\end{figure}

It is worth mentioning that the results of $R_{time}$, obtained from the simulation of supercritical networks, exhibit an exponential decay. For example, Fig.~\ref{fig:R_time} represents the results for $R_{time}$ in the supercritical network with density~$600$~nodes/$km^{2}$. As observed in this figure, the simulation results for $R_{time}$ exhibit an exponential decay, and the curve $a\cdot e^{-b\cdot t}$ fits the results well. Thus, functions $R_{time}(t)$ for supercritical networks can be estimated as
\begin{equation} \label{eq:estimatedR_time_super}
\centering
R_{time}(t)~\hat{=}~a\cdot e^{-b\cdot t},
\end{equation} where $a$ and $b$ are obtained from simulation. We have also tried fitting a curve to the simulation results of $R_{time}$ for subcritical networks, but it yields large errors. 
Table~\ref{tab:fittingLambdat} represents the values of parameters and error for fitted curves in Figs.~\ref{fig:Lambdat} and \ref{fig:R_time}.

\begin{table}

    \renewcommand{\arraystretch}{1.5}
	\setlength{\tabcolsep}{3pt}
	\begin{center}
	\caption{The values of parameters and the root mean square error for the fitted curves in Figs.~\ref{fig:Lambdat} and \ref{fig:R_time}.}
	\label{tab:fittingLambdat}
	\end{center}
	\centering
	\begin{tabular}{c|c|c|c|c}
		\hline
		\bfseries Figure & $\boldsymbol{a}$ & $\boldsymbol{b}$ & $\boldsymbol{c}$ & \bfseries RMSE\\

	     \hline
	     \hline
	     \bfseries \ref{fig:Lambdat}(a) & 6.602e-6 &   0.04008 & 1.975e-6 & 1.3306e-7\\
	    \hline
	   \bfseries  \ref{fig:Lambdat}(b) & 1.002e-5 & 0.1264 & 3.27e-6 & 5.2787e-7\\
	     \hline
	   \bfseries \ref{fig:R_time} &    730.7 &       0.1168 &  - & 23.6793\\
	     \hline
	     
	\end{tabular}
\end{table}



\subsection{Discussion}
As observed in Fig.~9, $\beta_{time}$ decreases as time increases. According to this figure, the approximate ranges of $\beta_{time}$ are $(2 \times 10^{-6}, 9\times 10^{-6})$ and $(3\times 10^{-6}, 15\times 10^{-6})$ when node density is 500 and 600 nodes/$km^2$, respectively. Note that the range corresponding to $\lambda=500$ is shorter than the range corresponding to $\lambda=600$. In fact, the range of $\beta_{time}$ becomes shorter as node density decreases such that $\beta_{time}$ converges to a constant when the network becomes sparse. This constant is the pairwise meeting rate since as represented in the ODE model given in (2), infection rate in a sparse network is computed by multiplying the pairwise meeting rate by $\overline{N}(t)\cdot (M-\overline{N}(t))$, which indicates that $\beta_{time}$ is equal to the pairwise meeting rate.
Similar to $\beta_{time}$, it can be concluded that $\beta_{num}$ eventually converges to the pairwise meeting rate as node density decreases. Thus, according to Eqs.~(11) and (15), parameter $b$ tends to zero as node density decreases. This can be observed in Table 2 where value of parameter $b$ for Fig.~9(a) is less than that for Fig.~9(b). Therefore, parameter $b$ shows how much the network is dense in terms of the number of nodes. It is worth mentioning that it is not reasonable to compare the values of parameter $b$ for the fitted curves in Fig 8 since the curve fitted in Fig.~8(a) has a power-law term whereas that fitted in Fig.~8(b) has an exponential term.

As it can be seen in Fig.~\ref{fig:R_time}, $R_{time}$ is a decreasing function of time in supercritical networks. Simulation results show that in supercritical networks, more than half of nodes become infected at time 0 excluding the transfer delays. For example, as mentioned in Section~5.1, the average number of infected nodes at time 0 in the networks with $\lambda=600$~nodes/$km^2$ and $L=5~km$ is around 7722. The reason is existence of a cluster containing a large fraction of
nodes with high probability \cite{haenggi2009stochastic}. If
the source belongs to this cluster, a large fraction of nodes
become infected at time 0. The decreasing behavior of the infection rate is also observed in sparse networks after half of nodes are infected. Note that the infection rate in a sparse network with $N$ infected nodes can be computed as \mbox{$N\cdot (M-N)\cdot \beta$} where $\beta$ is the pairwise meeting rate. Function $N\cdot (M-N)\cdot \beta$ is decreasing on the interval $(\lfloor M/2 \rfloor, M)$.

\subsection{ODE models}\label{sec:models}
By considering function $R_{num}(N)$ as an approximation for the infection rate as 
mentioned earlier, $\frac{d\overline{N}(t)}{dt}$ can be approximated by $\mathbb{E}_{t}(R_{num}(N))$ which equals \mbox{$\mathbb{E}_{t}(\beta_{num}(N)\cdot N\cdot (M-N))$} according to Eq.~(\ref{eq:betaN}). Note that $\mathbb{E}_{t}$ is the expectation over time, $t$, and variable $N$ depends on $t$. According to (\ref{eq:estimatedLambdaN_sub}) and (\ref{eq:estimatedLambdaN_super}), \mbox{$\mathbb{E}_{t}(\beta_{num}(N)\cdot N\cdot (M-N))$} is estimated by \mbox{$\mathbb{E}_{t}((a\cdot N^{-b}+c)\cdot N\cdot (M-N))$} and \mbox{$\mathbb{E}_{t}(a\cdot e^{-b\cdot N}\cdot N\cdot (M-N))$} for subcritical and supercritical networks, respectively. By considering \mbox{$(a\cdot N^{-b}+c)\cdot N\cdot (M-N)$} and \mbox{$a\cdot e^{-b\cdot N}\cdot N\cdot (M-N)$} as two functions of $N$ and interchanging expectation and these functions, \mbox{$\mathbb{E}_{t}((a\cdot N^{-b}+c)\cdot N\cdot (M-N))$} and \mbox{$\mathbb{E}_{t}(a\cdot e^{-b\cdot N}\cdot N\cdot (M-N))$} are approximated by \mbox{$(a\cdot\overline{N}(t)^{-b}+c)\cdot \overline{N}(t)\cdot (M-\overline{N}(t))$} and \mbox{$a\cdot e^{-b\cdot\overline{N}(t)}\cdot \overline{N}(t)\cdot (M-\overline{N}(t))$}, respectively. Note that $\mathbb{E}_{t}(N)$ is written as $\overline{N}(t)$. Thus, we propose the ODE models given in (\ref{eq:odeN2}) and (\ref{eq:odeN3}) for subcritical and supercritical DTNs, respectively.
\begin{eqnarray}
\label{eq:odeN2}
\frac{d\overline{N}(t)}{dt} =& (a\cdot\overline{N}(t)^{-b}+c)\cdot \overline{N}(t)\cdot(M-\overline{N}(t))\\
\label{eq:odeN3}
\frac{d\overline{N}(t)}{dt} =& a\cdot e^{-b\cdot \overline{N}(t)}\cdot \overline{N}(t)\cdot(M-\overline{N}(t))
\end{eqnarray}
It is worth mentioning that parameters $a$ and $b$ in each of Eqs.~(\ref{eq:odeN2}) and (\ref{eq:odeN3}) are determined separately from data.

Function $R_{time}(t)$ is an approximation for the infection rate at time $t$. According to (\ref{eq:estimatedR_time_super}), we propose the following ODE model for epidemic routing in supercritical networks.
\begin{equation} \label{eq:odet_super}
\centering
\frac{d\overline{N}(t)}{dt}= a\cdot e^{-b\cdot t}
\end{equation} According to (\ref{eq:odet_super}), 
$\overline{N}(t)$, $t>0$, can be computed as follows,
\begin{equation} \label{eq:N(t)_super}
\centering
\overline{N}(t)= -\frac{a}{b}\cdot(e^{-b\cdot t}-1)+\overline{N}(0).
\end{equation} (\ref{eq:N(t)_super}) indicates that the average number of infected nodes as a function of time exhibits exponential behavior. Based on (\ref{eq:N(t)_super}), the limit of $\overline{N}(t)$ as $t$ tends to infinity is,
\begin{equation}\label{eq:lim_N(t)}
\lim_{t\rightarrow\infty}\overline{N}(t)=\overline{N}(0)+\frac{a}{b}.
\end{equation}
On the other hand, $\lim_{t\rightarrow\infty}\overline{N}(t)$ should equal the total number of nodes, $M$. We use $\lim_{t\rightarrow\infty}\overline{N}(t)=M$ and (\ref{eq:lim_N(t)}) to derive a single-parameter model for epidemic routing in supercritical networks as follows. Based on \mbox{$\lim_{t\rightarrow\infty}\overline{N}(t)=M$} and (\ref{eq:lim_N(t)}), parameter $a$ can be computed as \mbox{$b\cdot(M-\overline{N}(0))$}. Thus, the performance of epidemic routing in supercritical networks can be modeled as
\begin{eqnarray}
\label{eq:odet_super2}
\frac{d\overline{N}(t)}{dt} =& b\cdot(M-\overline{N}(0))\cdot e^{-b\cdot t}\\
\overline{N}(t) =& (\overline{N}(0)-M)\cdot e^{-b\cdot t}+M
\end{eqnarray} where $b$ is obtained from simulation. It is worth mentioning that the proposed model given in (\ref{eq:odet_super2}) can be written in the following form which is very similar to the previous ODE model given in (\ref{eq:ode}) \cite{zhang2007performance}.
\begin{equation} \label{eq:odet_super3}
\centering
\frac{d\overline{N}(t)}{dt}= b\cdot(M-N(t)).
\end{equation}

According to (\ref{eq:betat}), \mbox{$R_{time}(t)=\beta_{time}(t)\cdot \overline{N}(t)\cdot (M-\overline{N}(t))$}. Thus, we propose the following ODE model for epidemic routing in non-sparse subcritical networks,
\begin{equation} \label{eq:odet1}
\centering
\frac{d\overline{N}(t)}{dt}= \beta_{time}(t)\cdot \overline{N}(t)\cdot(M-\overline{N}(t)).
\end{equation}
Using (\ref{eq:estimatedLambdat}), we rewrite Eq.~(\ref{eq:odet1}) as
\begin{equation} \label{eq:odet2}
\centering
\frac{d\overline{N}(t)}{dt}= (a\cdot e^{-b\cdot t}+c)\cdot \overline{N}(t)\cdot(M-\overline{N}(t)).
\end{equation}
(\ref{eq:odet2}) has an exponential term as (\ref{eq:odeN3}). (\ref{eq:odet2}) has the following closed form solution,
\begin{equation} \label{eq:N(t)_closed}
\centering
\overline{N}(t)= \frac{M\cdot e^{c\cdot M\cdot t}}{e^{\frac{a\cdot M\cdot e^{-b\cdot t}}{b}+k}+e^{c\cdot M\cdot t}}=\frac{M}{e^{\frac{a\cdot M\cdot e^{-b\cdot t}}{b}-c\cdot M\cdot t+k}+1},
\end{equation} where
\begin{equation} \label{eq:coefficinet}
\centering
k= \ln(M-\overline{N}(0))-\ln(\overline{N}(0))-\frac{a\cdot M}{b}.
\end{equation}

Once $\overline{N}(T)$ is obtained from the proposed ODE models, $p(T)=\frac{\overline{N}(T)-1}{M-1}$. Function $p(T)$ can be considered as cumulative distribution function of the delivery delay of epidemic routing when there is no deadline. In this case, the average delivery delay can be computed as $\int_{0}^{\infty} (1-p(t))dt$ \cite{spyropoulos2009routing}.

Parameters $a$, $b$, and $c$ of the model given in Eq. (\ref{eq:odet2}) can be obtained by fitting curve $a.e^{-b.t}+c$ to $\beta_{time}$. In addition to fitting curve to $\beta_{time}$, these parameters can be computed if values of $\overline{N}(t)$ are known at time 0 and three different positive times. Specifically, we show how parameters $a$, $b$, and $c$ of the model given in (\ref{eq:odet2}) can be computed if $\overline{N}(t)$ is obtained from the simulation for times 0, $t_1$, $t_2=2t_1$, and $t_3=3t_1$ where $t_1$ can be any arbitrary time ($t_1>0$). Given that $t_1$ can be chosen very small, we need only to simulate the network for a short interval of time, $[0, t_3]$, in order to obtain $\overline{N}(0)$, $\overline{N}(t_1)$, $\overline{N}(2t_1)$, and $\overline{N}(3t_1)$. Obtaining the parameters of the ODE models from the simulation is reasonable given that the parameter of the previous ODE model represented in Eq.~(\ref{eq:ode}), pairwise meeting rate, should be obtained from the simulation, too.


According to (\ref{eq:N(t)_closed}) and (\ref{eq:coefficinet}), we have
\begin{equation}
    e^{\frac{a\cdot M\cdot e^{-b\cdot t}}{b}-c\cdot M\cdot t+k}=\frac{M}{\overline{N}(t)}-1,
\end{equation}
\begin{equation}
    \frac{a\cdot M\cdot e^{-b\cdot t}}{b}-c\cdot M\cdot t+k=\ln(\frac{M}{\overline{N}(t)}-1),
\end{equation}
\begin{equation}
    \frac{a\cdot (e^{-b\cdot t}-1)}{b}-c\cdot t=M^{-1}\cdot\ln(\frac{\overline{N}(0)\cdot(M-\overline{N}(t))}{\overline{N}(t)\cdot(M-\overline{N}(0))}), 
\end{equation}
Let $f(t)$ and $z$ denote $M^{-1}\cdot\ln(\frac{\overline{N}(0)\cdot(M-\overline{N}(t))}{\overline{N}(t)\cdot(M-\overline{N}(0))})$ and $e^{-b\cdot t_1}$, respectively. Therefore,
\begin{equation}
    f(t_1)= \frac{a}{b}\cdot (e^{-b\cdot t_1}-1)-c\cdot t_1=\frac{a}{b}\cdot (z-1)-c\cdot t_1,
\end{equation}
\begin{equation}
f(2t_1)=\frac{a}{b}\cdot (e^{-2\cdot b\cdot t_1}-1)-2\cdot c\cdot t_1
=\frac{a}{b}\cdot (z^2-1)-2\cdot c\cdot t_1,
\end{equation}
\begin{equation}
    f(3t_1)=\frac{a}{b}\cdot (e^{-3\cdot b\cdot t_1}-1)-3\cdot c\cdot t_1
 = \frac{a}{b}\cdot (z^3-1)-3\cdot c\cdot t_1.
\end{equation}
In order to obtain parameters $a$, $b$, and $c$, we solve the following system of equations with the four variables $a$, $b$, $c$, and $z$.
\begin{equation}
    \frac{a}{b}\cdot (z-1)-c\cdot t_1=f(t_1)
\end{equation}
\begin{equation}
    \frac{a}{b}\cdot (z^2-1)-2\cdot c\cdot t_1=f(2t_1)
\end{equation}
\begin{equation}
    \frac{a}{b}\cdot (z^3-1)-3\cdot c\cdot t_1=f(3t_1)
\end{equation}

Let $y$ denote $\frac{a}{b}\cdot (z-1)$, then the system of equations can be written as follows,
\begin{equation}\label{eq:eq1}
    y-c\cdot t_1=f(t_1)
\end{equation}
\begin{equation}\label{eq:eq2}
    y\cdot (z+1)-2\cdot c\cdot t_1=f(2t_1)
\end{equation}
\begin{equation}\label{eq:eq3}
    y\cdot (z^2+z+1)-3\cdot c\cdot t_1=f(3t_1)
\end{equation}
Using (\ref{eq:eq2}), $z^2$ can be computed as follows,
\begin{equation}\label{eq:z2_1}
    z^2=(\frac{f(2t_1)+2\cdot c\cdot t_1}{y}-1)^2.
\end{equation}
On the other hand, subtracting (\ref{eq:eq2}) from (\ref{eq:eq3}) yields
\begin{equation}
    y\cdot z^2-c\cdot t_1=f(3t_1)-f(2t_1),
\end{equation} which results in
\begin{equation}\label{eq:z2_2}
    z^2=\frac{f(3t_1)-f(2t_1)+c\cdot t_1}{y}.
\end{equation}
According to (\ref{eq:z2_1}) and (\ref{eq:z2_2}), the following equality holds.
\begin{equation}\label{eq:z2_equality}
    (\frac{f(2t_1)+2\cdot c\cdot t_1}{y}-1)^2=\frac{f(3t_1)-f(2t_1)+c\cdot t_1}{y}
\end{equation}
According to (\ref{eq:eq1}), $y=f(t_1)+c\cdot t_1$. By substituting $y$ with $f(t_1)+c\cdot t_1$ in (\ref{eq:z2_equality}) and simplifying it, $c$ can be computed as follows,
\begin{equation}\label{eq:c}
    c=\frac{f(t_1)^2+f(2t_1)^2-f(t_1)\cdot( f(2t_1)+ f(3t_1))}{t_1\cdot (3\cdot f(t_1)-3\cdot f(2t_1)+f(3t_1))}.
\end{equation} Using Eqs.~(\ref{eq:eq1}), (\ref{eq:eq2}), and (\ref{eq:c}), it is straightforward to compute $y$ and $z$. Since $z=e^{-b\cdot t_1}$, $b=-\frac{\ln{z}}{t_1}$. Therefore, $b$ can be computed once $z$ is obtained. Finally, substituting $y$, $b$, and $z$ with their values in equation $y=\frac{a}{b}\cdot (z-1)$, we can compute variable $a$.

\section{Performance Evaluation}\label{sec:perf}

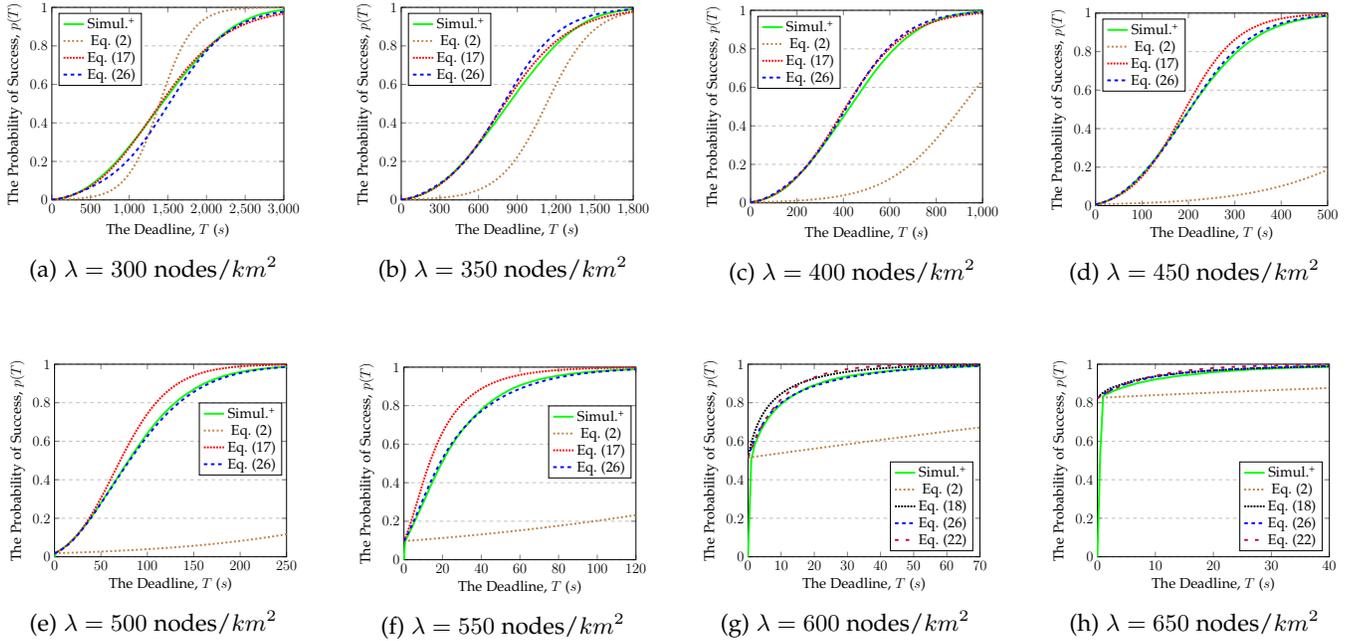
\begin{figure*}
\mbox{
\begin{subfigure}[b]{0.25\textwidth}
\centering
\begin{tikzpicture}[scale=0.45]
    \begin{axis}[
    xlabel={\large The Deadline, $T$ ($s$)},
    ylabel={\large The Probability of Success, $p(T)$},
    xmin=0, xmax=3000,
    ymin=0, ymax=1,
    xtick={0,500,1000,1500,2000,2500,3000},
    ytick={0,0.2,0.4,0.6,0.8,1},
    legend pos=north west,
    ymajorgrids=true,
    grid style=dashed,
     every axis plot/.append style={ultra thick},
     every tick label/.append style={font=\large}
    ]
     \addplot[
    color=green,
    mark=circle,
    mark options={solid},
    ] table [x=time, y=sp, col sep=space] {sp_300_less_points.txt};\addlegendentry{\large Simul.\textsuperscript{+}}
    
    
    \addplot[
    color=brown,
    mark=circle,
    mark options={solid},
    style=dotted
    ]
 table [x=time, y=sp, col sep=space] {earlyODE_300_less_points.txt};\addlegendentry{\large Eq.~(\ref{eq:ode})}
    
   \addplot[
    color=red,
    mark=circle,
    mark options={solid},
    style=densely dotted
    ]
 table [x=time, y=sp, col sep=space] {odeN_300.txt};\addlegendentry{\large Eq.~(\ref{eq:odeN2})}
 
    \addplot[
    color=blue,
  mark=circle,
    mark options={solid},
    style=dashed
    ]
 table [x=time, y=sp, col sep=space] {odet_300.txt};\addlegendentry{\large Eq.~(\ref{eq:odet2})}

    \end{axis}
\end{tikzpicture}
\caption{$\lambda=300$~nodes/$km^{2}$} 
\end{subfigure}
\begin{subfigure}[b]{0.25\textwidth}
\centering
\begin{tikzpicture}[scale=0.45]
    \begin{axis}[
    xlabel={\large The Deadline, $T$ ($s$)},
    ylabel={\large The Probability of Success, $p(T)$},
    xmin=0, xmax=1800,
    ymin=0, ymax=1,
    xtick={0,300,600,900,1200,1500,1800},
    ytick={0,0.2,0.4,0.6,0.8,1},
    legend pos=north west,
    ymajorgrids=true,
    grid style=dashed,
     every axis plot/.append style={ultra thick},
     every tick label/.append style={font=\large}
    ]
    
    \addplot[
    color=green,
    mark=circle,
    mark options={solid},
    ] table [x=time, y=sp, col sep=space] {sp_350_less_points.txt};\addlegendentry{\large Simul.\textsuperscript{+}}
    
    
     \addplot[
    color=brown,
    mark=circle,
    mark options={solid},
    style=dotted
    ]
 table [x=time, y=sp, col sep=space] {earlyODE_350.txt};\addlegendentry{\large Eq.~(\ref{eq:ode})}
    
    \addplot[
    color=red,
    mark=circle,
   mark options={solid},
    style=densely dotted
    ]
 table [x=time, y=sp, col sep=space] {odeN_350.txt};\addlegendentry{\large Eq.~(\ref{eq:odeN2})}
 
    \addplot[
    color=blue,
    mark=circle,
    mark options={solid},
    style=dashed
    ]
 table [x=time, y=sp, col sep=space] {odet_350.txt};\addlegendentry{\large Eq.~(\ref{eq:odet2})}

    \end{axis}
\end{tikzpicture}
\caption{$\lambda=350$~nodes/$km^{2}$} 
\end{subfigure}
\begin{subfigure}[b]{0.25\textwidth}
\centering
\begin{tikzpicture}[scale=0.45]
    \begin{axis}[
    xlabel={\large The Deadline, $T$ ($s$)},
    ylabel={\large The Probability of Success, $p(T)$},
    xmin=0, xmax=1000,
    ymin=0, ymax=1,
    xtick={0,200,400,600,800,1000},
    ytick={0,0.2,0.4,0.6,0.8,1},
    legend pos=north west,
    ymajorgrids=true,
    grid style=dashed,
     every axis plot/.append style={ultra thick},
     every tick label/.append style={font=\large}
    ]
    
    \addplot[
    color=green,
    mark=circle,
    mark options={solid},
    ] table [x=time, y=sp, col sep=space] {sp_400_less_points.txt};\addlegendentry{\large Simul.\textsuperscript{+}}
    
    
     \addplot[
    color=brown,
    mark=circle,
    mark options={solid},
    style=dotted
    ]
 table [x=time, y=sp, col sep=space] {earlyODE_400.txt};\addlegendentry{\large Eq.~(\ref{eq:ode})}
    
    \addplot[
    color=red,
    mark=circle,
    mark options={solid},
    style=densely dotted
    ]
 table [x=time, y=sp, col sep=space] {odeN_400.txt};\addlegendentry{\large Eq.~(\ref{eq:odeN2})}
 
    \addplot[
    color=blue,
    mark=circle,
    mark options={solid},
    style=dashed
    ]
 table [x=time, y=sp, col sep=space] {odet_400.txt};\addlegendentry{\large Eq.~(\ref{eq:odet2})}

    \end{axis}
\end{tikzpicture}
\caption{$\lambda=400$~nodes/$km^{2}$\\}

\end{subfigure}

\\

\begin{subfigure}[b]{0.25\textwidth}
\centering
\begin{tikzpicture}[scale=0.45]
    \begin{axis}[
    xlabel={\large The Deadline, $T$ ($s$)},
    ylabel={\large The Probability of Success, $p(T)$},
    xmin=0, xmax=500,
    ymin=0, ymax=1,
    xtick={0,100,200,300,400,500},
    ytick={0,0.2,0.4,0.6,0.8,1},
    legend pos=north west,
    ymajorgrids=true,
    grid style=dashed,
     every axis plot/.append style={ultra thick},
     every tick label/.append style={font=\large}
    ]
    
     \addplot[
    color=green,
    mark=circle,
    mark options={solid},
    ] table [x=time, y=sp, col sep=space] {sp_450_less_points.txt};\addlegendentry{\large Simul.\textsuperscript{+}}
    
 
  \addplot[
    color=brown,
    mark=circle,
    mark options={solid},
    style=dotted
    ]
 table [x=time, y=sp, col sep=space] {earlyODE_450.txt};\addlegendentry{\large Eq.~(\ref{eq:ode})}

 \addplot[
    color=red,
    mark=circle,
    mark options={solid},
    style=densely dotted
    ]
 table [x=time, y=sp, col sep=space] {odeN_450.txt};\addlegendentry{\large Eq.~(\ref{eq:odeN2})}
 
    \addplot[
    color=blue,
    mark=circle,
    mark options={solid},
    style=dashed
    ]
 table [x=time, y=sp, col sep=space] {odet_450.txt};\addlegendentry{\large Eq.~(\ref{eq:odet2})}

    \end{axis}
\end{tikzpicture}
\caption{$\lambda=450$~nodes/$km^{2}$\\}
\end{subfigure}

\\

}
\vspace*{0.05in}
\\

\mbox{

\begin{subfigure}[b]{0.25\textwidth}
\centering
\begin{tikzpicture}[scale=0.45]
    \begin{axis}[
    xlabel={\large The Deadline, $T$ ($s$)},
    ylabel={\large The Probability of Success, $p(T)$},
    xmin=0, xmax=250,
    ymin=0, ymax=1,
    xtick={0,50,100,150,200,250},
    ytick={0,0.2,0.4,0.6,0.8,1},
    legend style={at={(0.98,0.8)}},
    ymajorgrids=true,
    grid style=dashed,
     every axis plot/.append style={ultra thick},
     every tick label/.append style={font=\large}
    ]
    
    \addplot[
    color=green,
    mark=circle,
    mark options={solid},
    ] table [x=time, y=sp, col sep=space] {sp_500_less_points.txt};\addlegendentry{\large Simul.\textsuperscript{+}}
    
 
  \addplot[
    color=brown,
    mark=circle,
    mark options={solid},
    style=dotted
    ]
 table [x=time, y=ODE, col sep=space] {earlyODE_500.txt};\addlegendentry{\large Eq.~(\ref{eq:ode})}
    
    \addplot[
    color=red,
    mark=circle,
    mark options={solid},
    style=densely dotted
    ]
 table [x=time, y=sp, col sep=space] {odeN_500.txt};\addlegendentry{\large Eq.~(\ref{eq:odeN2})}
 
    \addplot[
    color=blue,
    mark=circle,
    mark options={solid},
    style=dashed
    ]
 table [x=time, y=sp, col sep=space] {odet_500.txt};\addlegendentry{\large Eq.~(\ref{eq:odet2})}
 

    \end{axis}
\end{tikzpicture}
\caption{$\lambda=500$~nodes/$km^{2}$} 
\end{subfigure}
\begin{subfigure}[b]{0.25\textwidth}
\centering
\begin{tikzpicture}[scale=0.45]
    \begin{axis}[
    xlabel={\large The Deadline, $T$ ($s$)},
    ylabel={\large The Probability of Success, $p(T)$},
    xmin=0, xmax=120,
    ymin=0, ymax=1,
    xtick={0,20,40,60,80,100,120},
    ytick={0,0.2,0.4,0.6,0.8,1},
    legend style={at={(0.98,0.8)}},
    ymajorgrids=true,
    grid style=dashed,
     every axis plot/.append style={ultra thick},
     every tick label/.append style={font=\large}
    ]
    
    \addplot[
    color=green,
    mark=circle,
    mark options={solid},
    ] table [x=time, y=simulPosDelay, col sep=space] {comparison.txt};\addlegendentry{\large Simul.\textsuperscript{+}}
    
    \addplot[
    color=brown,
    mark=circle,
    mark options={solid},
    style=dotted
    ]
 table [x=time, y=meetingRate, col sep=space] {comparison.txt};\addlegendentry{\large Eq.~(\ref{eq:ode})}

    \addplot[
    color=red,
    mark=circle,
    mark options={solid},
    style=densely dotted
    ]
 table [x=time, y=sp, col sep=space] {odeN_550.txt};\addlegendentry{\large Eq.~(\ref{eq:odeN2})}
 
    \addplot[
    color=blue,
    mark=circle,
    mark options={solid},
    style=dashed
    ]
 table [x=time, y=sp, col sep=space] {odet_550.txt};\addlegendentry{\large Eq.~(\ref{eq:odet2})}

    \end{axis}
\end{tikzpicture}
\caption{$\lambda=550$~nodes/$km^{2}$\\} 
\end{subfigure}

\\

\begin{subfigure}[b]{0.25\textwidth}
\centering
\begin{tikzpicture}[scale=0.45]
    \begin{axis}[
    xlabel={\large The Deadline, $T$ ($s$)},
    ylabel={\large The Probability of Success, $p(T)$},
    xmin=0, xmax=70,
    ymin=0, ymax=1,
    xtick={0,10,20,30,40,50,60,70},
    ytick={0,0.2,0.4,0.6,0.8,1},
    legend pos=south east,
    ymajorgrids=true,
    grid style=dashed,
     every axis plot/.append style={ultra thick},
     every tick label/.append style={font=\large}
    ]
    
    \addplot[
    color=green,
    mark=circle,
    mark options={solid},
    ] table [x=time, y=sp, col sep=space] {sp_600_less_points.txt};\addlegendentry{\large Simul.\textsuperscript{+}}
    
    \addplot[
    color=brown,
    mark=circle,
    mark options={solid},
    style=dotted
    ]
 table [x=time, y=earlyODE, col sep=space] {sp_earlyODE_600_less_points.txt};\addlegendentry{\large Eq.~(\ref{eq:ode})}
    
    \addplot[
    color=black,
    mark=circle,
    mark options={solid},
    style=densely dotted
    ]
 table [x=time, y=sp, col sep=space] {odeN_600.txt};\addlegendentry{\large Eq.~(\ref{eq:odeN3})}
 
    \addplot[
    color=blue,
    mark=circle,
    mark options={solid},
    style=dashed
    ]
 table [x=time, y=sp, col sep=space] {odet_600.txt};\addlegendentry{\large Eq.~(\ref{eq:odet2})}
 
  \addplot[
    color=purple,
    mark=circle,
    mark options={solid},
    style=loosely dashed
    ]
 table [x=time, y=sp, col sep=space] {odet_super_oneVariable_600.txt};\addlegendentry{\large Eq.~(\ref{eq:odet_super2})}
 

    \end{axis}
\end{tikzpicture}
\caption{$\lambda=600$~nodes/$km^{2}$} 

\end{subfigure}
\begin{subfigure}[b]{0.25\textwidth}
\centering
\begin{tikzpicture}[scale=0.45]
    \begin{axis}[
    xlabel={\large The Deadline, $T$ ($s$)},
    ylabel={\large The Probability of Success, $p(T)$},
    xmin=0, xmax=40,
    ymin=0, ymax=1,
    xtick={0,10,20,30,40},
    ytick={0,0.2,0.4,0.6,0.8,1},
    legend pos=south east,
    ymajorgrids=true,
    grid style=dashed,
     every axis plot/.append style={ultra thick},
     every tick label/.append style={font=\large}
    ]
    
    \addplot[
    color=green,
    mark=circle,
    mark options={solid},
    ] table [x=time, y=sp, col sep=space] {sp_650_less_points.txt};\addlegendentry{\large Simul.\textsuperscript{+}}
    
    \addplot[
    color=brown,
    mark=circle,
    mark options={solid},
    style=dotted
    ]
 table [x=time, y=earlyODE, col sep=space] {earlyODE_650.txt};\addlegendentry{\large Eq.~(\ref{eq:ode})}
    
    \addplot[
      color=black,
      mark=circle,
    mark options={solid},
    style=densely dotted
    ]
 table [x=time, y=sp, col sep=space] {odeN_650.txt};\addlegendentry{\large Eq.~(\ref{eq:odeN3})}
 
    \addplot[
    color=blue,
   mark=circle,
    mark options={solid},
    style=dashed
    ]
 table [x=time, y=sp, col sep=space] {odet_650.txt};\addlegendentry{\large Eq.~(\ref{eq:odet2})}

  \addplot[
    color=purple,
    mark=circle,
    mark options={solid},
    style=loosely dashed
    ]
 table [x=time, y=sp, col sep=space] {sp_oneVariable_650.txt};\addlegendentry{\large Eq.~(\ref{eq:odet_super2})}

    \end{axis}
\end{tikzpicture}
\caption{$\lambda=650$~nodes/$km^{2}$}
\end{subfigure}
}
\caption{The probability of success computed from the ODE models given in Eqs.~(\ref{eq:ode}), (\ref{eq:odeN2}), (\ref{eq:odeN3}), (\ref{eq:odet_super2}), and (\ref{eq:odet2}) and the results obtained from the simulation.}
\label{fig:ODESimul}
\end{figure*}

In this section, we validate models presented in this paper. Fig.~\ref{fig:ODESimul} presents $p(T)$ obtained by the proposed ODE models given in Eqs.~(\ref{eq:odeN2}), (\ref{eq:odeN3}), (\ref{eq:odet_super2}), and (\ref{eq:odet2}), the original ODE model given in Eq.~(\ref{eq:ode}), and simulation for a network with $L=5~km$, \mbox{$R=50~m$}, \mbox{$v_{min}=0$}, \mbox{$v_{max}=1~m/s$}, and \mbox{$t_{r}=120~s$} considering different densities. Transfer rate of each connection and message size are taken to be 50~\textit{Mbps} and 25 \textit{KB}, respectively, as in Section \ref{sec:pairwiseRate}. Figs.~\ref{fig:ODESimul}(a), \ref{fig:ODESimul}(b), \ref{fig:ODESimul}(c), \ref{fig:ODESimul}(d), and \ref{fig:ODESimul}(e) present results for subcritical networks. The density of the network corresponding to Fig.~\ref{fig:ODESimul}(f) is just below the critical density. Furthermore, Figs.~\ref{fig:ODESimul}(g) and \ref{fig:ODESimul}(h) present results for supercritical networks.



Excluding the estimate of $p(T)$ computed from the original ODE model in (\ref{eq:ode}), a ramp-up is observed in each curve presented in Figs.~\ref{fig:ODESimul}(a), \ref{fig:ODESimul}(b), \ref{fig:ODESimul}(c), \ref{fig:ODESimul}(d), and \ref{fig:ODESimul}(e) unlike curves presented in Figs.~\ref{fig:ODESimul}(f), \ref{fig:ODESimul}(g), and \ref{fig:ODESimul}(h). When node density is close to the percolation critical density, or it is higher than the percolation critical density, large clusters consisting of a considerable fraction of nodes can form with non-negligible probability. If the source initially belongs to such a cluster, and transfer delays are neglected, the source infects a large fraction of nodes, belonging to that cluster, at time 0 avoiding the ramp-ups observed in Figs.~\ref{fig:ODESimul}(a), \ref{fig:ODESimul}(b), \dots, and \ref{fig:ODESimul}(e). When transfer delays are greater than zero, no node becomes infected at time 0 by the source, and $p(0)=0$ as it can be easily observed in curves \textit{Simul.\textsuperscript{+}} presented in Fig.~\ref{fig:ODESimul}(f), \ref{fig:ODESimul}(g), and \ref{fig:ODESimul}(h). Since transfer delays are neglected by the proposed ODE models and the original ODE model, analytic results presented in Fig.~\ref{fig:ODESimul} are not zero at $T=0$. Although simulation results are zero at $T=0$ in Figs.~\ref{fig:ODESimul}(f), \ref{fig:ODESimul}(g), and \ref{fig:ODESimul}(h), they exhibit abrupt changes when $T$ is close to 0 representing that the source infects a large fraction of nodes in the cluster to which it initially belongs after a short time.



As observed in Fig.~\ref{fig:ODESimul}, the proposed models given in Eqs.~(\ref{eq:odeN3}), (\ref{eq:odet_super2}), and (\ref{eq:odet2}) are very accurate. Furthermore, the ODE model given in Eq.~(\ref{eq:odeN2}) is accurate enough to predict $p(T)$ under epidemic routing in subcritical networks except that node density is 500 or 550. On the other hand, the original ODE model given in~(\ref{eq:ode}) cannot predict $p(T)$, and yields a significant error at all node densities as discussed in Section~\ref{sec:pairwiseRate}. This indicates the superiority of the proposed ODE models in comparison with the ODE model proposed in~\cite{zhang2007performance}. In conclusion, the ODE model given in (\ref{eq:odet2}) that uses pairwise infection rate as a function of
time is preferred over the ODE models given in (\ref{eq:odeN2}), (\ref{eq:odeN3}), and (\ref{eq:odet_super2}) since it can be used to evaluate the performance of both subcritical and supercritical networks, and yields good accuracy at all node densities.

One can think that ODE models given in Eqs.~(\ref{eq:odeN2}) and (\ref{eq:odeN3}) do not work so well because, as mentioned in Section~\ref{sec:betaN}, $\beta_{num}$ exhibits neither a power low behavior nor an exponential behavior when the number of infected nodes is close to the total number of nodes. However, it is not the reason since $p(T)$ is close to one when the number of infected nodes is close to the total number of nodes, and ODE models given in Eqs.~(\ref{eq:odeN2}) and (\ref{eq:odeN3}) have good accuracy when $p(T)$ is near one as observed in Fig.~\ref{fig:ODESimul}. For example, as mentioned in Section~\ref{sec:betaN}, behavior of $\beta_{num}$ in the intervals (12000, 12500) and (14000, 15000) respectively for densities 500 and 600 is excluded. When the number of infected nodes belongs to these intervals, $p(T)$ is more than 0.93. As it can be seen in Figs.~\ref{fig:ODESimul}(e) and \ref{fig:ODESimul}(g), ODE models given in Eqs.~(\ref{eq:odeN2}) and (\ref{eq:odeN3}) have good accuracy when $p(T)$ is more than 0.93.

As mentioned in Section~\ref{sec:models}, the ODE models given in Eqs.~(\ref{eq:odeN2}) and (\ref{eq:odeN3}) are derived based on interchanging \mbox{$(a\cdot N^{-b}+c)\cdot N\cdot (M-N)$} and \mbox{$a\cdot e^{-b\cdot N}\cdot N\cdot (M-N)$} as two functions of $N$ and expectation over time, $t$. This interchange is a major simplification. Probably, the reason the ODE models given in Eqs.~(\ref{eq:odeN2}) and (\ref{eq:odeN3}) are weaker than the ODE model given in Eq.~(\ref{eq:odet2}) is this interchange. As mentioned in Section~\ref{sec:pairwiseRate}, the infection rate depends on many factors such as the number of infected nodes and time. In the ODE model given in Eq.~(\ref{eq:odet2}), the infection rate is considered as a function of the time and the average number of infected node at that time while it is a function of only time in the model given in Eq.~(\ref{eq:odet_super2}). This difference might be the reason that the model given in Eq.~(\ref{eq:odet2}) is more accurate than the model given in Eq.~(\ref{eq:odet_super2}) as it can be seen in Figs.~\ref{fig:ODESimul}(g) and \ref{fig:ODESimul}(h). Another alternative for reason is the difference between the two models in terms of the number of parameters. The model given in Eq.~(\ref{eq:odet_super2}) has only one parameter while the model given in (\ref{eq:odet2}) has three parameters. Moreover, Table \ref{tab:fittingLambdat} represents that for the same network setting, the RMSE for fitting curve to $R_{time}$ is much more than the RMSE for fitting curve to $\beta_{time}$.




\section{Conclusion and future work} \label{sec:conc}
The goal of this paper was to analyze the performance of epidemic routing in non-sparse DTNs. We first proved that PS experiences a phase transition as node density increases. This result was validated by simulation of large finite networks. Afterwards, we showed a shortcoming of the standard ODE model \cite{zhang2007performance} in modeling non-sparse networks.
In order to overcome with this difficulty, we proposed four new ODE models to evaluate the performance of epidemic routing in non-sparse DTNs. Two models were proposed for subcritical and supercritical networks, respectively, and use pairwise infection rate as a function of the number of infected nodes. On the other hand, we proposed a model that works for both subcritical and supercritical networks, and uses pairwise infection rate as a function of time. Moreover, we derived a single-parameter model to evaluate the performance of epidemic routing in supercritical networks based on infection rate as a function of time.
Comparison of results computed from the proposed ODE models and the results obtained from the simulation showed that the models proposed for supercritical networks and the model using pairwise infection rate as a function of time are accurate. Moreover, the other model, proposed for subcritical networks based on pairwise infection rate as a function of the number of infected nodes, is accurate when density is far from the percolation critical density.

As future work, it is a good idea to investigate whether or not a phase transition happens in the performance of other well-known routing schemes such as \textit{K}-hop forwarding. In addition to PS, it is worth studying some other important performance measures such as delivery delay.
Another future research direction can be proposing ODE models for other routing schemes in non-sparse DTNs.


%



\ifCLASSOPTIONcaptionsoff
  \newpage
\fi



%
\bibliographystyle{IEEEtran}
\bibliography{IEEEabrv,Myreferences}

%

\begin{IEEEbiography}[{\includegraphics[width=1in,height=1.25in,clip,keepaspectratio]{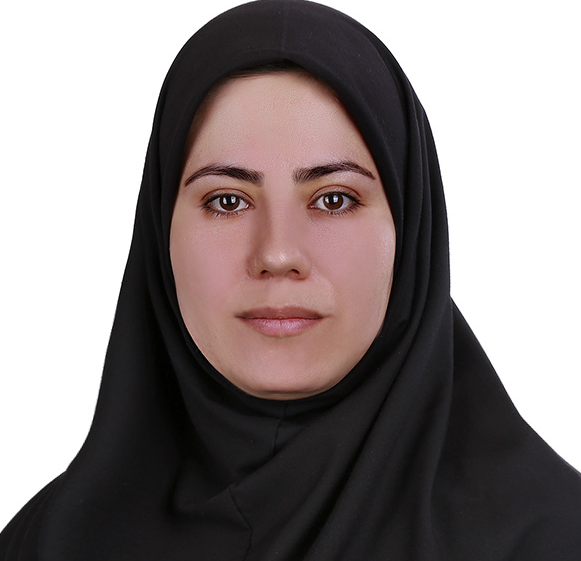}}]{Leila Rashidi} is a postdoctoral associate at the Department of Computer Science, University of Calgary, Calgary, Canada. She received her Ph.D. in Computer Engineering from the Department
of Computer Engineering, Sharif University of Technology, Iran in 2019, and the B.S. degree in Computer Engineering from the University of Tehran,
Tehran, Iran, in 2014. She was a visiting researcher at University of
Massachusetts Amherst and Imperial College London in 2017 and 2018, respectively. Her main research interests
are performance evaluation, mobile networks, optimization, and network security.
\end{IEEEbiography}

\begin{IEEEbiography}[{\includegraphics[width=1in,height=1.25in,clip,keepaspectratio]{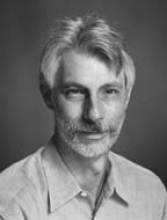}}]{Don Towsley} received the B.A. degree in physics and the Ph.D. degree in computer science from the University of Texas in 1971 and 1975, respectively. From 1976 to 1985, he was a Faculty Member of the Department of Electrical and Computer Engineering, University of Massachusetts, Amherst, MA, USA. He is currently a Distinguished Professor with the College of Information and Computer Sciences, University of Massachusetts. He has held visiting positions at the IBM T. J. Watson Research Center, Yorktown Heights, NY, USA; Laboratoire MASI, Paris, France; INRIA, Sophia-Antipolis, France; AT\&T Labs-Research, Florham Park, NJ, USA; and Microsoft Research Lab, Cambridge, U.K. His research interests include networks and performance evaluation. He has been elected as a Fellow of the ACM. He is a member of the ORSA, and the Chair of IFIP Working Group 7.3. He has received the 2007 IEEE Koji Kobayashi Award, the 2007 ACM SIGMETRICS Achievement Award, the 1998 IEEE Communications Society William Bennett Best Paper Award, and numerous best conference/workshop paper awards. He was the Program Co-Chair of the joint ACM SIGMETRICS and PERFORMANCE 92 Conference, and the Performance 2002 Conference. He served as the Editor-in-Chief for the IEEE/ACM Transactions on Networking, and he also serves on the Editorial Board for the Journal of the ACM and the IEEE Journal on Selected Areas in Communications. He has previously served on numerous other editorial boards.
\end{IEEEbiography}

\begin{IEEEbiography}[{\includegraphics[width=1in,height=1.25in,clip,keepaspectratio]{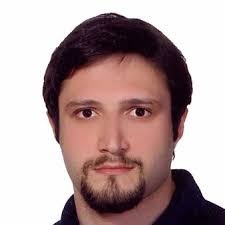}}]{Arman Mohseni-Kabir} is a physics Ph.D. candidate at the University of Massachusetts Amherst. His research interests include network science and statistical physics with a focus on percolation theory and machine learning on graph structures. Arman received his master’s degree in physics from Sharif University of Technology in 2013.
\end{IEEEbiography}



\begin{IEEEbiography}[{\includegraphics[width=1in,height=1.25in,clip,keepaspectratio]{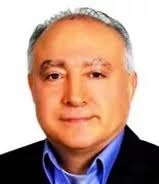}}]{Ali Movaghar} is a Professor in the Department of Computer Engineering at Sharif University of Technology. He received his B.S. degree in Electrical Engineering from the University of Tehran in 1977, and M.S. and Ph.D. degrees in Computer, Information, and Control Engineering from the University of Michigan, in 1979 and 1985, respectively.
He visited the Institut National de Recherche en Informatique et en Automatique in Paris, France and the Department of Electrical Engineering and Computer Science at the University of California, Irvine in 1984 and 2011, respectively, worked at AT\&T Information Systems in Naperville, IL in 1985-1986, and taught at the University of Michigan, Ann Arbor in 1987-1989.
His research interests include performance/dependability modeling and formal verification of wireless networks and distributed real-time systems.
\end{IEEEbiography}




\end{document}